\documentclass[a4paper,11pt,fleqn]{article}
\usepackage[left=2.50cm, right=2.50cm, top=3.50cm, bottom=2.00cm]{geometry}
\usepackage[utf8]{inputenc}
\usepackage{amsthm}
\usepackage{color}
\usepackage[fleqn]{amsmath}
\usepackage{graphicx}
\usepackage{amssymb}
\usepackage{url}
\usepackage{hyperref}
\usepackage{todonotes}
\usepackage{verbatim}

\makeatother
\newcommand{\cmt}[1]{\todo[color=green!15!white, linecolor=black]{#1}}
\newcommand{\cmtjj}[1]{\iffalse #1 \fi}
\newcommand{\cmti}[1]{\todo[color=green!15!white,linecolor=black,inline]{#1}}
\newcommand{\cmtii}[1]{\iffalse #1 \fi}

\newcommand{\tsblue}{\color{blue!80!green!80!black}}
\newcommand{\tsgrn}{\color{black}}
\newcommand{\cbk}{\color{black}}
\newcommand{\rd}{\,{\rm d}}
\newcommand{\scrT}{\mathcal{T}} 
\newcommand{\bmac}[0]{\frac{s}{l^2}}
\title{Quasi-local instantaneous charges asymptotics at spatial infinity}
\author{Jacek Jezierski\thanks{\texttt{Jacek.Jezierski@fuw.edu.pl}}\; and Tomasz Smołka\thanks{Corresponding author: \texttt{Tomasz.Smolka@fuw.edu.pl}}  \\
	Department of Mathematical Methods in
	Physics, \\ University of Warsaw,
	Pasteura 5, 02-093 Warszawa, Poland}
\newtheorem{defi}{Definition}[section]
\newtheorem{theorem}{Theorem}[section]
\newtheorem{prop}{Proposition}[section]
\newtheorem{lemat}{Lemma}[section]
\begin{document}
	\numberwithin{equation}{section}
	\maketitle
\begin{abstract}
	The article aims to analyze a construction of charges (conserved quantities) for the gravity field in the (3+1) decomposition. The construction is based on (3+1) splitting of conformal Yano--Killing (CYK) two-form. The splitting leads to charges, defined on Cauchy surface, which are combined from components of Weyl tensor and conformal Killing vector. The relations between the conserved quantities and its classical ADM counterparts are revisited. Asymptotic behavior of the conserved quantities is described. The charges are analyzed for a particular choice of initial data, among others, Bowen -- York spinning black hole.
\end{abstract}
\section{Introduction}

 Conserved quantities play a key role in description of gravitating systems. 
 The issue of energy that can be radiated by a system in de Sitter universe is a question of current interest. This problem has been addressed by Ashtekar et al. in~\cite{ashtekar2014asymptotics}. It has been revisited later from different perspective, for example by Chruściel and Ifsits~\cite{ChIfsits}. The analysis by Chruściel, Jezierski and Kijowski~\cite{chrusciel2015hamiltonian} showed\footnote{See introduction and appendices in the cited paper.}\cmtjj{Doprecyzowac referencje?} a proper choice of Hamiltonian field, including overall factor and suitable three-dimensional Cauchy surface is not fully clear in the case of de Sitter spacetimes. The question then arises, whether some insight into the problem at hand could be gained from different perspective---by establishing a relation between a conserved quantity for non-linear gravity, constructed from four-dimensional objects: conformal Yano--Killing (CYK) two-forms and (linearized) Weyl tensor, and quasi-local instantaneous charges defined on three-dimensional Cauchy surface. Such charges have been introduced by Ashtekar and Hansen in \cite{ashtekarhansen1978unified}. We note that the proposed approach to conserved quantities, based on the relation between CYK two-forms and conformal Killing vectors (CKV), gives a more general perspective on the issue of choosing CKV and Cauchy surface in a more compatible way. To be precise,  we analyze the relation between \eqref{eq:def_I}, and quasi-local instantaneous charges, defined by \eqref{eq:def_Q_E} and \eqref{eq:def_Q_B}.
 The investigations of spacetimes with positive cosmological constant (see\footnote{The citation \cite{ashtekar2014asymptotics} is the first paper from a series of papers published by the same authors.} \cite{ashtekar2014asymptotics}), show the charges \eqref{eq:def_Q_E} and \eqref{eq:def_Q_B} enables one to avoid interpretation problems related to casual properties of de Sitter asymptotic regime. In particular, mass in ADM approach is defined with the help of time translation generator which changes its causal nature when passing through cosmological horizon. Mass defined as an instantaneous charge requires an existence of scaling generator on a three-dimensional Cauchy surface. Using scaling generator instead of time translation generator enables one to avoid interpretation problems considering mass in de Sitter spacetimes. Moreover, we recall a classical result given by Ashtekar which states that the ADM angular momentum depends on supertranslations (for example see \cite{ashtekar1984supertranslations} and the references within). In other words, the ADM formula for angular momentum is coordinate dependent. The supertranslation ambiguities can be removed by imposing stronger boundary conditions at spatial infinity (see \cite{regge1974role}). One of us has shown in \cite{Jezierski1995RelationSPinmetric} that the existence of asymptotic CYK tensor $Q_{ACYK}$ which fulfills the CYK equation, see \eqref{eq:CYK_eq1} and \eqref{eq:CYK_def}, asymptotically\footnote{The equation \eqref{eq:ACYK} mean all the components of ${\cal Q}_{\lambda \kappa
         \sigma}(Q_{ACYK})$ fall off like $r^{-1}$ or faster. $r$ is a radial coordinate which is well-defined in asymptotically flat regime.} 
 \begin{equation}
 {\cal Q}_{\lambda \kappa
     \sigma}(Q_{ACYK}) \approx r^{-1} \, ,
 \label{eq:ACYK}
 \end{equation}
 removes the supertranslational ambiguity. See appendix \ref{sec:Strong_asymp_ACYK} for details.

Our investigations are initially performed for Minkowski spacetime and conformally flat spacetimes. Part of statements is further generalized to de Sitter spacetimes. (CYK) two-forms are related to so called \emph{hidden symmetries} of spacetime. The readers familiar with CYK two-forms may worry about a possible generalization of our approach for curved spacetimes. Preliminary analysis in this context in~\cite{czajkajezierski} leads to a conclusion that the results in this paper carries over without major difficulties to de Sitter background.

We hope that our idea would bring a new air into a problem of conserved
charges for asymptotically de Sitter spacetimes. A meaningful notion of
energy should respect finite transfer of energy from the gravitational
field to other physical fields.  Additionally, we plan to examine
conditions which would guarantee that data on the Cauchy (or null)
hypersurface is physically reasonable.
One can consider spacetimes in which a space of CYK symmetries is not
maximal. In this case it is worth to analyze the asymptotic charges.
For this, we will analyze properties of the ACYK solution (see appendix
\ref{sec:Strong_asymp_ACYK}).
 
\noindent 
 \textbf{Overview of this paper.} Used conventions and denotings are listed in the Appendix (section \ref{sec:notations}). In section \ref{sec:CYK_survey}, we give a brief survey on CYK two-forms. CYK two-form is a generalization of conformal Killing covector to skew-symmetric p-forms. These geometrical objects, associated with so called \textit{hidden symmetries}, significantly simplify a description of electromagnetism or weak gravitational perturbation. CYK forms also enable one to construct charges. For details see \cite{jezsmo,jezierski2015_3_decomp}.  Section \ref{sec:Instant_charges} is devoted to (3+1) decomposition of CYK tensors. In which theorem \ref{thm:charge_decomp} about (3+1) charge splitting is one of the key points of our construction. The relations between ADM linear momentum and angular momentum and corresponding instantaneous charges are presented in section \ref{sec:Linear_angular momentum}. The results agrees with the other proposals to define a notion of quasi-local quantities associated with a closed 2-surface (see, e.g., \cite{penrose1982quasi}).
 
  We provide definitions and discuss basic properties of
 gravitational charges, or quantities specified on the spatial hypersurface
 $\Sigma_t$ immersed in four-dimensional spacetime.
 The structure of quasi-local charges presented here can be used for
 any initial data $(\gamma _{ij}, K _{ij})$, for which
 three-dimensional metric $\gamma _{ij}$ is conformally flat 
 or for the three-metric which approaches conformally flat metric at infinity.
  The presented construction is illustrated by a particular choice of examples in section \ref{sec:examples}. Comparison of conditions which are required for the asymptotic conservation of ADM quantities and instantaneous charges is given in section \ref{sec:asymptotic_charges}.

\section{Survey on CYK tensors and associated charges \label{sec:CYK_survey}}
Let $Q_{\mu\nu}$ be a skew-symmetric tensor field (two-form) on a four-dimensional manifold $M$ and let us denote by ${\cal Q}_{\lambda \kappa \sigma}$ a
(three-index) tensor which is defined as follows:
\begin{equation}
\label{eq:CYK_eq1}
{\cal Q}_{\lambda \kappa
	\sigma}(Q,g):= Q_{\lambda \kappa ;\sigma} +Q_{\sigma \kappa
	;\lambda} - \frac{2}{3} \left( g_{\sigma
	\lambda}Q^{\nu}{_{\kappa ;\nu}} + g_{\kappa (\lambda }
Q_{\sigma)}{^{\mu}}{_{ ;\mu}} \right) \, .
\end{equation}
The object $\cal Q$ has the following algebraic properties
\begin{equation}
\label{wlQ}
{\cal Q}_{\lambda\kappa\mu}g^{\lambda\mu}=0=
{\cal Q}_{\lambda\kappa\mu}g^{\lambda\kappa} \, , \quad
{\cal Q}_{\lambda\kappa\mu} =
{\cal Q}_{\mu\kappa\lambda}\, ,
\end{equation}
i.e. it is traceless and partially symmetric.
\begin{defi}\label{CYK_df}
	A skew-symmetric tensor $Q_{\mu\nu}$ is a conformal Yano--Killing tensor
	(or simply CYK tensor) for the metric $g$ iff
	\begin{equation}
	\label{eq:CYK_def}
	{\cal Q}_{\lambda \kappa \sigma}(Q,g) = 0
	\end{equation}
\end{defi}		
CYK tensors are generalization of conformal Killing vectors to two-forms.  The first proposition of such object is given in \cite{tachibana1969conformal}. Properties and detailed information can be found in \cite{jezierski2006cykkerr} and in the references within. The set of equations in definition \ref{CYK_df} is overdetermined. Four dimensional, conformally flat spacetimes admit maximal (20 dimensional) space of CYK tensor solutions. Some solutions exist for type D spacetimes \cite{walker1970quadratic}. In particular, generalized Plebański--Demiański black hole contains at least a pair of solutions \cite{kubizvnak2007CYK_PD}. In the context of further research is important the following property
\begin{theorem}[Hodge duality]\label{dual_th}
	Let $g_{\mu\nu}$ be a metric of a four-dimensional differential
	manifold $M$. $*$ denotes Hodge duality\footnote{Hodge duality (Hodge star) is given by $*\omega_{\alpha\beta} :=
		\frac{1}{2}\varepsilon_{\alpha\beta}{}^{\mu\nu}\omega_{\mu\nu} \, ,$
		where $\varepsilon_{\alpha\beta\mu\nu}$ is the skew-symmetric
		Levi--Civita tensor.}. A skew-symmetric tensor $Q_{\mu\nu}$ is a CYK tensor
	of the metric $g_{\mu\nu}$ iff its dual $*Q_{\mu\nu}$ is a CYK
	tensor of this metric.
\end{theorem}
The theorem has been proved in \cite{jezierski2006cykkerr}. See also the generalized form of the statement in \cite{yasui2011hidden}.   
The conserved quantities are constructed from the spin-2 tensor field. \tsgrn  It is convenient to recall properties of spin-2 tensor field: \cbk
\begin{eqnarray}
    W_{\alpha\beta\mu\nu} = W_{\mu\nu\alpha\beta} = W_{[\alpha\beta][\mu\nu]} \, , \quad W_{\alpha[\beta\mu\nu]} = 0\, , \quad g^{\alpha\mu}W_{\alpha\beta\mu\nu}=0 \, , \quad \nabla_{[\lambda} W_{\mu\nu ] \alpha\beta} =0 \, .
\end{eqnarray}
The following equality holds
\begin{equation}
\nabla_{[\lambda}W_{\mu\nu]\alpha\beta} = 0 \iff \nabla^{\mu}W_{\mu\nu\alpha\beta} = 0 \, .
\end{equation}
$W_{\mu\nu\alpha\beta}$ is skew-symmetric both in the first and
the second pair of indices and to both of them the Hodge star can
be applied. We denote
\begin{equation} \label{eq:hodge_spin_2}
{}^{*}W_{\mu\nu\alpha\beta} = \frac{1}{2}\varepsilon_{\mu\nu}{}^{\rho\sigma}
W_{\rho\sigma\alpha\beta} \ ,  \quad W^{*}{}_{\mu\nu\alpha\beta} =
\frac{1}{2}W_{\mu\nu\rho\sigma}\varepsilon^{\rho\sigma}{}_{\alpha\beta}\ .
\end{equation}
The symmetries of spin-2 tensor provides: ${}^{*}W = W^{*}, \; {}^{*}({}^{*}W) = {}^{*}W^{*} = - W. \,$ Let $W_{\mu\nu\alpha\beta}$ be a spin-2 field and $Q_{\mu\nu}$
be a CYK tensor. Let us denote by $F$ the following two-form:
\begin{equation}\label{def_FWQ}
F_{\mu\nu}(W,Q) := W_{\mu\nu\lambda\kappa}Q^{\lambda\kappa}.
\end{equation}
The following formula is satisfied:
\begin{equation}\label{divF}
\nabla_{\nu}F^{\mu\nu}(W,Q) =0 \, .
\end{equation}
Theorem \ref{dual_th} and Eq. (\ref{eq:hodge_spin_2}) enable one to observe similarly
\begin{eqnarray} \label{eq:F_duality}
*F_{\mu \nu}= {}^{*}W_{\mu\nu\lambda\kappa}Q^{\lambda\kappa}=W_{\mu\nu\lambda\kappa}(*Q)^{\lambda\kappa} \, , \quad \nabla_{\nu}(*F)^{\mu\nu}(W,Q) =0 \, .
\end{eqnarray}
Let $V$ be a three-volume and $\partial V$ its boundary. Formula
(\ref{divF}) implies\footnote{Symbols $\rd \sigma_{\mu\nu}$
	and $\rd \Sigma_{\mu}$
	can be defined in the following way: if $\Omega$ stands for the volume
	form of the manifold $M$, then $\rd \sigma_{\mu\nu}:=(\partial_\mu
	\wedge \partial_\nu) \lrcorner\, \Omega$, $\rd \Sigma_{\mu}:=
	\partial_\mu \lrcorner\, \Omega$.}
\[
\int_{\partial V} F^{\mu\nu}(W,Q) \rd \sigma_{\mu\nu} = \int_{V}
\nabla_{\nu} F^{\mu\nu}(W,Q) \rd \Sigma_{\mu} = 0 \, .
\]
In this sense $Q_{\mu\nu}$ defines a charge related to the spin-2
field $W$\footnote{The flux of $F^{\mu\nu}$ through any
	two-dimensional closed surfaces $S_{1}$ and $S_{2}$ is the same as
	long as we are able to find a three-volume $V$ between them (i.e. there
	exists $V$ such that $\partial V = S_{1} \cup S_{2}$). }:
\begin{equation} \label{eq:def_I}
C(W,Q):=\int_{S} F^{\mu\nu}(W,Q) \rd \sigma_{\mu\nu} \, .
\end{equation}
\tsgrn where $S$ is a two-dimensional closed surface. \cbk 
\section{(3+1) decomposition of CYK tensor for Minkowski spacetime. Instantaneous charges \label{sec:Instant_charges}}
The aim of this section is to present the decomposition of CYK tensor and associated conserved quantities for Minkowski spacetime. A brief supplement about used properties of conformal Killing vector field is contained in the appendix \ref{sec:CKV}.  
\subsection{(3+1) decomposition of CYK tensor}
  In \cite{jezierski2015_3_decomp}, the following decomposition of CYK tensor for Minkowski spacetime has been proved
\begin{lemat} \label{lemat2} Each CYK tensor in Minkowski spacetime can be expressed in the following way:\begin{equation}
     Q=a(t) \scrT_0 \wedge X + b(t) *\!(\scrT_0 \wedge
	Y), \label{obserwacja} \end{equation} 
    where $X$, $Y$ are (three-dimensional)
	conformal Killing fields; $a(t)$, $b(t)$
	are  quadratic polynomials of a single indeterminate $t$.
\end{lemat}
The basis of the space of solutions for the
equations, given in definition \ref{CYK_df} (i.e. the basis of CYK
tensors) in Minkowski spacetime is twenty-dimensional (see \cite{houri2015simple,houri2018integrability}). This is a maximal possible dimension of space of CYK solutions for four-dimensional spacetime.\\
\tsgrn
For Minkowski spacetime, each CYK two-form is a linear combination of a wedge product of two conformal Killing co-vectors (CKV) or a Hodge dual of such product, given by \eqref{obserwacja}. This makes the (3+1) decomposition of CYK tensor simple. In particular, the choice of Cauchy surface $t=\mathrm{const}$ is natural.  In the case of de Sitter spacetime or even curved spacetimes, the splitting of CYK form is much more complicated and the choice of suitable Cauchy surface is much less intuitive. However, the CYK tensor decomposition should guarantee existence of Gaussian charges on a properly chosen surface. Deep understanding of the construction in the flat case is required for further analysis of de Sitter spacetimes.
\\
\cbk
Spin-2 tensor can be splitted into well-known gravitoelectromagnetic tensors\footnote{Let $n^\mu$ be a normed vector perpendicular to the foliation $\Sigma_t$. A spin-2 field on $\Sigma_{t}$ can be equivalently described by two symmetric and traceless tensors: the
	electric and the magnetic part of Weyl tensor,
	\begin{equation}
	E_{\alpha\beta}:= W_{\alpha\mu\nu\beta}n^\mu n^\nu\,, \quad B_{\alpha\beta}:= W^*{}_{\alpha\mu\nu\beta}n^\mu n^\nu\,.
	\end{equation}}. 
Let us consider how the charge (\ref{eq:def_I}) is related with instantaneous charges (\ref{eq:def_Q_E}), (\ref{eq:def_Q_B}) on Cauchy surface $\Sigma_t: t=\mathrm{const}$. We need to introduce the following conserved quantity:
\begin{defi}[\tsgrn instantaneous charges \cbk] \label{def:instantaneous_charges}
	Consider a given spatial hypersurface $\Sigma$ equipped with a Riemannian metric $\gamma$. Let $A$ be a two-dimensional closed surface, embedded in $\Sigma$. We can define the following instantaneous charges:
	\begin{eqnarray}
	I(E,X)&:=& \int_A E ^{i}{}_{j}X^j dS_i\,, \label{eq:def_Q_E}\\
	I(B,X)&:=& \int_A B ^{i}{}_{j}X^j dS_i\,, \label{eq:def_Q_B}
	\end{eqnarray}
	where $X^{j}$ is a conformal Killing vector field (CKV). 
\end{defi} 
 The relation between four-dimensional CYK conserved quantity and CYK charge is as follows. Using (\ref{eq:F_duality}), (\ref{eq:def_I}), (\ref{obserwacja}), and decomposition into magnetic and electric part we can prove 
\begin{theorem} \label{thm:charge_decomp}
	Let $\Sigma_{t}$ be a $t=\mathrm{const}$ Cauchy surface in Minkowski spacetime. Charge  $C(W,Q)$, defined by (\ref{eq:def_I}), can be decomposed at each $\Sigma_{t}$ surface as follows
	\begin{equation}
    \label{eq:thm_CYK_decomposition}
	C(W,Q)= \alpha(t)I(E,X)+ \beta(t) I(B,Y) \, , 
	\end{equation} 
	where $I(E,X)$ and $I(B,Y)$  are instantaneous charges (\ref{eq:def_Q_E}) and (\ref{eq:def_Q_B}) respectively; $\alpha(t)$, $\beta(t)$
	are quadratic polynomials of a single indeterminate $t$.   
\end{theorem}
 The equation (\ref{eq:thm_CYK_decomposition})indicate time dependence of $C(W,Q)\, .$ It can be clearly seen for linearized gravity, see pages 9-10 in \cite{jezierski2015_3_decomp}. In non linear theory, the issue is much more complicated. Such time dependences can be observed in Plebański--Demiański family of black holes \cite{griffiths2006new}, in particular C-metric is the simplest case in which such time dependences occurs.
The charges defined by (\ref{eq:def_Q_E}) and (\ref{eq:def_Q_B}) will not depend on the choice of a two-dimensional surface if the appropriate divergence is zero. The charges have been proposed by Ashtekar and Hansen in \cite{ashtekarhansen1978unified}. If we consider two two-dimensional, oriented closed surfaces $A_1$, $A_2$ limiting the three-dimensional volume $V: \partial V = A_1 \cup A_2$, then the conformal Killing vector $X^i$ and the electrical part $E _{ij}$ fulfill the following equation:
\begin{equation}
\label{powierzchnieS}
\int_{A_1} \sqrt{\gamma} E ^{i}{}_{j}X^j dS_i - \int_{A_2} \sqrt{\gamma} E
^{i}{}_{j}X^j dS_i = \int_V (\sqrt{\gamma} E ^{i}{}_{j}X^j)_{,i}dV\,.
\end{equation}
The divergence of vector density $\sqrt{\gamma}E ^{i}{}_{j}X^j$ can be decomposed as follows: 
\begin{equation}
\label{divergencecalculation}
\begin{aligned}
(\sqrt{\gamma}E ^{i}{}_{j}X^j) _{,i} &= 	(\sqrt{\gamma}E ^{i}{}_{j}X^j) _{|i} =
\sqrt{\gamma}E ^{i}{}_{j|i}X^j + \sqrt{\gamma} E ^{ij}X _{j|i} = \\
&=\sqrt{\gamma} E ^{i}{}_{j|i}X^j + \sqrt{\gamma} E ^{ij}X _{(j|i)} =
\sqrt{\gamma} E ^{i}{}_{j|i}X^j + \sqrt{\gamma} E ^{ij} \frac{\lambda}{2}g _{ij} =\\
&=\sqrt{\gamma} E ^{i}{}_{j|i}X^j\,,
\end{aligned}
\end{equation}
where $E _{ij}$ is symmetric and traceless, and $X^i$ is a CKV.
It enables one to formulate the following proposition
\begin{prop} \label{prop:IC_conservation}
	Consider a three-dimensional Cauchy surface with a Riemannian metric $\gamma$. Let $S$ be a two-dimensional topological sphere embedded in volume $V$. If $X$ be a conformal Killing vector for metric $\gamma$ and the gravitoelectric tensor field is divergenceless on $V$  then the instantaneous charge
	\begin{equation*}
	I(E,X)=\int_A E ^{i}{}_{j}X^j dS_i\,,
	\end{equation*}
	does not depend on the choice of integration surface $A \subset V$. The analogue result holds for $I(B,X)$, defined by (\ref{eq:def_Q_B}).   
\end{prop}
Ambiguity of choice of a surface $A$ can be understood as follows:
Let us consider a topological ball $\mathcal{B}$ inside the three-dimensional spatial space $\sigma$. In $\mathcal{B}$, we have $(\sqrt{\gamma}E ^{i}{}_{j}X^j) _{,i}\neq0$ in contrary to outside region $\Sigma \setminus \mathcal{B}$ where $(\sqrt{\gamma}E ^{i}{}_{j}X^j) _{,i}=0$. Using Stokes Theorem, one obtain 
\begin{equation}
\int_{\partial \mathcal{B}} \sqrt{\gamma}E ^{i}{}_{j}X^j=\int_{A}\sqrt{\gamma}E ^{i}{}_{j}X^j
\end{equation}
where $A$ is a topological sphere which surrounds $\mathcal{B}$. $A$ may be any surface with the above property.

\subsection{Gravitoelectromagnetic tensors in terms of initial data}
In \cite{2} the electromagnetic parts of Weyl tensor are expressed in terms of initial data $(\gamma _{ij},K _{ij})$ on the Cauchy surface $\Sigma$. The results are the following:
\begin{theorem}
	\label{theoremEBlambda}
	If spacetime fulfills the Einstein vacuum equations with a
	cosmological constant:\begin{equation}\label{tw2Einsteinvaceq}
	R _{\mu \nu}-\frac{1}{2}R g _{\mu \nu}+\Lambda g _{\mu\nu}=0\,,
 	\end{equation}
	then the electrical and magnetic parts of the Weyl tensor are expressed by the initial data $(\gamma _{ij}, K _{ij})$ on the three-dimensional spatial hypersurface
	$\Sigma$ as follows: \begin{eqnarray}
	\label{lematElambda}
	E _{i j} &=& -\overset{3}{R}{}_{i j} - K K _{i j} +
	K _{i k}K ^{k}{}_{j}+\frac{2}{3}\Lambda \gamma_{ij}\,, \\
	\label{lematBlambda}
	B _{ij} &=& \varepsilon ^{ls}{}_{j}D_s K _{il}\,,
	\end{eqnarray}
	where $\overset{3}{R}{}_{ij}$ is a Ricci tensor of a three-dimensional metric, $K
	_{ij}$ is a tensor of extrinsic curvature, and  $K=K ^{i}{}_{i}$ is a trace of extrinsic curvature.
\end{theorem}
 \noindent The proof is mainly based on $(3+1)$ decomposition of Weyl tensor and geometrical identities between four-dimensional and three-dimensional curvature tensors.
Similar decomposition can be done for the divergence of the electric (magnetic) part of Weyl tensor
\begin{theorem}
	\label{divergenceTheorem}
	If spacetime fulfills the Einstein vacuum equations with a cosmological constant, then the three-dimensional covariant divergence of the electrical part $E$
	and the magnetic part $B$ of Weyl tensor is expressed as follows:
	\begin{eqnarray}
	E ^{i}{}_{j|i} &=& (K \wedge B)_{j}\,, \label{eq:div_E_KwB}\\
	B ^{i}{}_{j|i} &=& -(K \wedge E) _{j}\,, \label{eq:div_B_Kwe}
	\end{eqnarray}
	where $\wedge$ is an operation defined for two symmetric tensors $A$ and
	$B$ as
	\begin{equation}
	(A \wedge B)_a :=
	\varepsilon _{a}{} ^{bc}A _{b}{}^dB _{dc}\,.
	\end{equation}
\end{theorem}
\noindent The theorem simplifies examination of conservancy\footnote{See the discussion below (\ref{divergencecalculation}).} of the charges (\ref{eq:def_Q_E}) and (\ref{eq:def_Q_B}). It can be especially useful in situations, where  the ``pure'' electrical (with zero magnetic part) or ``pure'' magnetic (with zero electrical part) is analyzed. 
\section{ADM mass, ADM linear momentum, ADM angular momentum and corresponding instantaneous charges \label{sec:Linear_angular momentum}}
\subsection{Physical interpretation of instantaneous charges} \label{sec:physical_interpretation}
Let us observe the relations between traditional quantities (e.g. ADM or Komar formula) and instantaneous charges. More precisely, we have the following table:\\
\begin{table}[h!]
\centerline{
	\begin{tabular}{|c|c|c|l|}
		\hline
		KV & CKV& Charge & Description \\
		\hline
		$\mathcal{T}_0$ &$\mathcal{S}$& $I\left(E,\mathcal{S}\right)$  & energy (mass) \\
		$\mathcal{T}_k$ &$\mathcal{R}[k]$& $I\left(B,\mathcal{R}[k]\right)$&  linear momentum \\
		$\mathcal{L}_{kl}$&$\mathcal{K}[k]$& $I\left(B,\mathcal{K}[k]\right)$  & angular momentum \\
		$\mathcal{L}_{0k}$ &$\mathcal{K}[k]$& $I\left(E,\mathcal{K}[k]\right)$  & center of mass \\
		\hline
	\end{tabular}
}
\caption{ \tsgrn First column contain an ADM/Komar generator of a charge. In the second column a corresponding instantaneous conserved quantity generator is given. In the last one a physical interpretation of a charge is provided.}
\end{table}\\
Other quantities: dual mass, dual momentum,
linear acceleration, and angular acceleration
are usually vanishing.
However, some parameters in Einstein metrics can be interpreted as topological charges, e.g. dual mass appears in Taub-NUT
solution (\cite{JJMLnut},\cite{jezierski2007asymptotic}) and dual momentum in Demia\'nski metrics (\cite{Jezierski1995RelationSPinmetric},\cite{griffiths2006new}).
We discuss below the relation between ADM quantities and instantaneous charges.
\subsubsection{Mass}
Mass is one of the most important quantity which characterize a physical system. However, some subtleties arise and they are analyzed in section \ref{sec:examples}. We claim that the following instantaneous charge
\begin{equation} \label{eq:mass_def}
\mathcal{M}:=\frac{1}{8 \pi}I(E,\mathcal{S})
\end{equation}
describes quasi-local mass. $\mathcal{S}$ is CKV related with scaling generator\footnote{For conformally flat space, with the metric in the form $\psi\left( \rd R^2+ R^2 \rd \Omega^2 \right)$ the scaling generator has a form $\mathcal{S}=R \partial_{R} \, .$}. If we consider asymptotically flat space with two-dimensional foliation of topological spheres $S(R)$, which are parametrized by radius $R$ then the ADM mass is given by $M_{ADM}= \frac{1}{8 \pi}\lim\limits_{R \to \infty} \mathcal{M}(S(R))$.

\subsubsection{Linear momentum}
\begin{theorem}
	\label{theoremmomentum}
	Let $\Sigma$ be a flat, three-dimensional spatial hypersurface immersed in the spacetime which satisfies Einstein vacuum equations.
	Assuming that the three-dimensional covariant divergence of the magnetic part disappears:
	\begin{equation}
	\label{zalozenieodywergencejiliniowy}
	B ^{i}{}_{j|i}=0\,,
	\end{equation}
	then for any two-dimensional, closed surface $A$ immersed in
	$\Sigma$ holds:
	\begin{equation}
	\label{tezaLiniowy}
	\int_A B ^{i}{}_{j} \mathcal{R} ^{j}[k]dS_i =
	\int_A P ^{i}{}_{j} \mathcal{T} ^{j}[k]dS_i\,,
	\end{equation}
	where $P ^{i}{}_{j}$ is (canonical) ADM momentum, $\mathcal{R}[k]$ is a rotation generator around the axis
	$k$, and $\mathcal{T}[k]$ is a translation generator along the axis $k$.
\end{theorem}
\noindent\textit{Proof:}\\
The surface $\Sigma$ is assumed to be flat, so appropriate conformal Killing fields exist. \\
First we show that an integral over any surface $A$ can be converted into an integral over a two-dimensional sphere.\\
The ADM momentum is expressed by the extrinsic curvature: 
\begin{equation}
P ^{i}{}_{j} = -K ^{i}{}_{j} + K \gamma ^{i}{}_{j}\,.
\end{equation}
Einstein vacuum equations are satisfied, so in particular vacuum vector constraint:
\begin{equation}
\label{lemmaconst}
K ^{i}{}_{j|i}-K _{|j}=0 \quad \Rightarrow \quad P ^{i}{}_{j|i}=0	\,.
\end{equation}
The divergence in the integral in the equation (\ref{tezaLiniowy}) can be reformulated analogically to the equation (\ref{divergencecalculation}). Assuming
(\ref{zalozenieodywergencejiliniowy}), we obtain that the divergence of the term containing the magnetic part is zero. For the integral on the right hand side of the equation (\ref{tezaLiniowy}), we have:
\begin{equation}
(\sqrt{\gamma} P ^{i}{}_{j} \mathcal{T}^{j}[k]) _{,i}=
(\sqrt{\gamma} P ^{i}{}_{j} \mathcal{T}^{j}[k]) _{|i}=
\sqrt{\gamma} P ^{i}{}_{j|i}\mathcal{T}^j[k] + \sqrt{\gamma}P ^{(ij)}  (\mathcal{T}[k])
{}_{(j|i)}=0\,.
\end{equation}
 After applying the Leibniz rule, the first term in the above equation is zero from the equation (\ref{lemmaconst}), and in the second term we can add symmetrization, the translation generator fulfills the Killing equation, hence the symmetrized covariant derivative of $\mathcal{T}$ vanishes. \\
We have shown that the divergences of both integrands in equation (\ref{tezaLiniowy}) are zero, using the Stokes theorem, the volume term has no contribution, which
proves formulae (\ref{tezaLiniowy}). \tsgrn Without loss of generality, we can set $k=z$. \cbk Replacing the integral on
any surface $A$ by an integral on a two-dimensional sphere $S(r)$ is justified.\\
The left hand side of (\ref{tezaLiniowy}) is:
\begin{equation}
\int_{S(r)} B ^{i}{}_{j} \mathcal{R} ^j_z dS_i = \int_{S(r)} \lambda B
^{r}{}_{\phi}\,,
\end{equation}
where $\lambda = r^2\sin\theta$. Using (\ref{tezaLiniowy}), we have
\begin{eqnarray}
B ^{r}{}_{\phi}&=& \varepsilon ^{rij} K _{\phi i |j} \nonumber \\
&=&
\varepsilon ^{rAB} K _{\phi A | B} r^2 \varepsilon ^{\phi
	C}(\cos\theta) _{,C} \nonumber \\
&=&
\varepsilon ^{AB}K _{DA|B}r^2 \varepsilon ^{DC} (\cos\theta)
_{,C}\, ,
\end{eqnarray}
where the convention for two-dimensional Levi--Civita tensor is $r^2\sin\theta \varepsilon
^{\theta\phi}=1$. A three-dimensional covariant derivative can be decomposed into:
\begin{equation}
K _{DA|B} = K _{DA,B} - \Gamma ^{m}{}_{DB}K _{mA} - \Gamma ^{m}{}_{AB}K
_{Dm}= K _{DA||B} + \frac{1}{r}\eta _{AB}K _{Dr}+\frac{1}{r}\eta _{DB}K
_{rA}\,,
\end{equation}
where we used the fact that $\Gamma^{r}{_{AB}} = - \frac{1}{r}\eta _{AB}$,
true for a covariant derivative of a flat three-dimensional space, and $\eta
_{AB}$ is a metric induced on a sphere with the radius $r$, it gives:
\begin{eqnarray}
B ^{r}{}_{\phi} &=&
\varepsilon ^{AB}K _{DA||B}r^2 \varepsilon ^{DC} (\cos\theta)
_{,C}+r \varepsilon ^{A}{}_{D}K _{rA} \varepsilon ^{DC} (\cos\theta) _{,C} \nonumber \\
&=&
\varepsilon ^{AB}K _{DA||B}r^2 \varepsilon ^{DC} (\cos\theta)
_{,C}+ \frac{1}{r} K _{r\theta}  \sin\theta\, ,
\end{eqnarray}
where the identity $\varepsilon ^{A}{}_{D}\varepsilon ^{DC}= -\eta ^{AC}$ holds.
\begin{equation}
\label{linearmomentumlemma}
\begin{aligned}
\int_{S(r)} \lambda B ^{r}{}_{\phi} &=
\int_{S(r)}\lambda\varepsilon ^{AB}K _{DA||B}r^2 \varepsilon ^{DC} (\cos\theta)
_{,C}+ \int_{S(r)}\lambda \frac{1}{r} K _{r\theta}  \sin\theta = \\
=&-\int_{S(r)}\lambda\varepsilon ^{AB}K _{DA}r^2 \varepsilon ^{DC} (\cos\theta)
_{||CB}+ \int_{S(r)}\lambda \frac{1}{r} K _{r\theta}  \sin\theta = \\
=&\int_{S(r)}\lambda\varepsilon ^{AB}K _{DA} \varepsilon ^{DC}
\eta _{CB}\cos\theta+ \int_{S(r)}\lambda \frac{1}{r} K _{r\theta}  \sin\theta = \\
=&\int_{S(r)}\lambda K _{DA}  \eta^{DA}
\cos\theta+ \int_{S(r)}\lambda \frac{1}{r} K _{r\theta}  \sin\theta \,,
\end{aligned}
\end{equation}
where we have used the identity $r^2(\cos\theta)
_{||CB} = -\eta _{CB}\cos\theta$.\\
The field $\mathcal{T}_z$ in the spherical coordinates:
\begin{equation}
\mathcal{T}[z] = \partial_z = \cos\theta \partial_r -
\frac{\sin\theta}{r}\partial_\theta\,.
\end{equation}
Therefore, the right hand side of the thesis (\ref{tezaLiniowy}) takes the form:
\begin{equation}
\begin{aligned}
-\int_{S(r)} P ^{i}{}_{j} \mathcal{T}^j[z] dS_i &= \int_{S(r)} \left( K
^{i}{}_{j} - \delta ^{i}{}_{j} K \right) \mathcal{T}^j [z] dS_i &=& \\
&= \int_{S(r)} \lambda (K ^{r}{}_{r}-K)\cos\theta -\lambda
\frac{1}{r}\sin\theta K ^{r}{}_{\theta} &=& \\
&= \int_{S(r)}  -K ^{AB}\eta _{AB}\lambda\cos\theta -\lambda
\frac{1}{r}\sin\theta K {}_{r\theta} &=&\quad - \int_{S(r)} \lambda B
^{r}{}_{\phi}\,,
\end{aligned}
\end{equation}
where the last equality comes from the comparison with the right hand side of
(\ref{linearmomentumlemma}).
\qed
\subsubsection{Angular momentum}
\begin{theorem}
	\label{theoremangularmomentum}
	Let $\Sigma$ be a flat, three-dimensional spatial hypersurface immersed in the spacetime which satisfies Einstein vacuum equations. Assuming that the following charges vanish:
	\begin{equation}
	\label{zerowezwezenia}
	I(B,\mathcal{T}[x]) = I(B, \mathcal{T}[y]) = I(B, \mathcal{T}[z])=0
	\end{equation}
	and the three-dimensional covariant divergence of the magnetic part disappears:
	\begin{equation}
	\label{zalozenieodywergenceji}
	B ^{i}{}_{j|i}=0\,,
	\end{equation}
	then for any two-dimensional, closed surface $A$ immersed in
	$\Sigma$ holds:
	\begin{equation}
	\label{teza}
	\int_A B ^{i}{}_{j} \mathcal{K} ^{j}[k]dS_i =
	\int_A P ^{i}{}_{j} \mathcal{R} ^{j}[k]dS_i\,,
	\end{equation}
	where $P ^{i}{}_{j}$ is the ADM momentum, $\mathcal{K}[k]$ is the generator of proper conformal transformations in the direction of $k$, and
	$\mathcal{R}[k]$ is a rotation generator around the axis $k$.
\end{theorem}
\noindent\textit{Proof:}\\
Analogically to the proof of theorem \ref{theoremmomentum}, it can be shown that the divergences of integrands in the thesis (\ref{teza})
disappear, that justifies the proof for integrals on
two-dimensional spheres. \\
By assumption (\ref{zerowezwezenia}), we can deduce that the integral from the contraction of the magnetic part with any vector with constant coefficients in the
Cartesian system is zero (because $\mathcal{T}[k] = \partial_k$). Using
this observation, we have:
\begin{equation}
\label{stalywektor}
\int_{S(r)} B ^{i}{}_{j}( \mathcal{K}^j[k] + A^j)dS_i = 	
\int_{S(r)} B ^{i}{}_{j} \mathcal{K}^j[k] dS_i  	\,,
\end{equation}
where $A^j$ are the coordinates of a constant vector in the Cartesian system. Any
such a vector can be written in a spherical system in the following way:
\begin{equation}
A^i \partial_i = \frac{A_ix^i}{r}\partial_r + r \left(
\frac{A_i x^i}{r}
\right)^{,B}\partial_B\,.
\end{equation}
We denote $u:= (A_ix^i)/r$, then:
\begin{equation}
A^i \partial_i = u\partial_r + r (u)^{,B}\partial_B\,.
\end{equation}
Now we choose $u$ to simplify the calculation of the integral (\ref{stalywektor}). The generators of the conformal transformations in the spherical system read:\cmtjj{Change $\mathcal{K}_{j}->\mathcal{K}[j]$}
\begin{eqnarray}
\mathcal{K}[x] &=& \frac{1}{2}r^2\cos\phi\sin\theta\partial_r -
\frac{1}{2}r\cos\phi\cos\theta\partial_\theta+ \frac{1}{2}r
\frac{\sin\phi}{\sin\theta}\partial_\phi\,, \\
\mathcal{K}[y] &=& \frac{1}{2}r^2\sin\phi\sin\theta\partial_r -
\frac{1}{2}r\sin\phi\cos\theta\partial_\theta - \frac{1}{2}r
\frac{\cos\phi}{\sin\theta}\partial_\phi\,,\\
\mathcal{K}[z] &=&
\frac{1}{2}r^2\cos\theta\partial_r+\frac{1}{2}r\sin\theta\partial_\theta\,,
\end{eqnarray}
Let us choose the function $u$:\\
For $\mathcal{K}[x]$ we choose $u[x]=\frac{1}{2}r^2\cos\phi\sin\theta$.\\
For $\mathcal{K}[y]$ we choose $u[y]=\frac{1}{2}r^2\sin\phi\sin\theta$.\\
For $\mathcal{K}[z]$ we choose $u[z]=\frac{1}{2}r^2\cos\theta$.\\
We will denote new ``corrected'' vectors by $\bar{\mathcal{K}}[k] :=
\mathcal{K}[k] + (A[k])^i \partial_i$, and we obtain:
\begin{equation}
\bar{\mathcal{K}}[x] = r^2\cos\phi\sin\theta\partial_r = r^2 \frac{\partial
	r}{\partial x}\partial_r\,,
\end{equation}
\begin{equation}
\bar{\mathcal{K}}[y] = r^2\sin\phi\sin\theta\partial_r = r^2 \frac{\partial
	r}{\partial y}\partial_r\,,
\end{equation}
\begin{equation}
\bar{\mathcal{K}}[z] = r^2\cos\theta\partial_r= r^2 \frac{\partial
	r}{\partial z}\partial_r\,.
\end{equation}
The vectors $\bar{\mathcal{K}}[k]$ now have only the component in the radial direction.\\
Without loss of generality we consider (\ref{teza}) for $k$
equal to $z$. Coordinate system that the $z$ axis is turned in the direction which is invariant under the rotation generator $\mathcal{R}[k]$ always can be chosen.\\
In the spherical system, the metric on $\Sigma$
takes the form:
\begin{equation}
\gamma _{ij} =  dr^2 + r^2 d\theta^2 + r^2\sin^2\theta d\phi^2\,,
\end{equation}
and the metric induced on the spheres:
\begin{equation}
\eta_{AB} =  r^2d\theta^2 + r^2\sin^2\theta d\phi^2 \,.
\end{equation}
Note that the rotation generator field around the $z$ axis such that $\mathcal{R}[z]=\partial_\phi$ can be written as:
\begin{equation}
\partial_\phi =  r^2 \varepsilon ^{AB} \left( \cos\theta
\right)_{,B}\partial_A\,,
\end{equation}
where $\varepsilon ^{AB}$ is an antisymmetric tensor on the sphere, by definition:
\begin{equation}
\sqrt{\eta} \varepsilon ^{\theta\phi}=  r^2 \sin\theta \varepsilon
^{\theta\phi}=1\,.
\end{equation}
We denote: $\lambda = \sqrt{\gamma}$ and reformulate the
right hand side of (\ref{teza}):
\begin{equation}
\label{prawamoment}
\begin{aligned}
\int\limits_{S(r)}P ^{i}{}_{j} \mathcal{R}^j[z] dS_i
&=
-\int\limits_{S(r)} \lambda
K ^{r}{}_{A}   r^2 \varepsilon ^{AB}
(\cos\theta) _{,B}
=\int\limits_{S(r)} r^2(\lambda
K_{rA}    \varepsilon ^{AB})_{||B}
\cos\theta = \\ &
=\int\limits_{S(r)} r^2\lambda
\varepsilon ^{AB}K_{rA||B}
\cos\theta \,.
\end{aligned}
\end{equation}

For the left hand side of the thesis:
\begin{equation}
\label{lewamoment}
\begin{aligned}
\int_{S(r)} B ^{i}{}_{j}\mathcal{K}^j[z] dS_i &= \int_{S(r)} \lambda B ^{r}{}_{j}
\bar{\mathcal{K}}^j[z] = \int_{S(r)} \lambda \varepsilon ^{lsr}K _{jl|s}
\bar{\mathcal{K}}^j[z] = \int_{S(r)} \lambda \varepsilon ^{ABr} K _{rA|B}
r^2\cos\theta = \\ &= \int_{S(r)} r^2\lambda \varepsilon ^{AB} K
_{rA||B}\cos\theta\,.
\end{aligned}
\end{equation}
By comparing the formulae (\ref{prawamoment}) and (\ref{lewamoment}) we get the thesis.
\qed 
\section{Examples}
\label{sec:examples}
We apply the concepts introduced in section \ref{sec:Instant_charges} to the analysis of conserved quantities for particular choices of initial data. The aim of this section is to perform a detailed discussion of two of the most important parameters which characterizes a black hole solution: mass and angular momentum. We compare our results with classical ADM approach and present methods of obtaining quasi-local quantities (like quasi-local mass). 
\tsgrn
First two examples contain initial data on particularly chosen surfaces which can be considered in Schwarzschild, or in Schwarzschild--de Sitter, or in Schwarzschild--Anti de Sitter spacetimes. \cbk We will consider two families of
foliations with spatial hypersurfaces $\Sigma$. The first of them, denoted
$\Sigma_s$, corresponds to the surfaces of constant time (i.e. the coordinate appearing in the standard form of Schwarzschild metric)
$t=\textrm{const}$, the second (denoted $\Sigma_p$) will be a hypersurface foliation with a flat inner geometry.
Next example belongs to Bowen--York initial data type. We discuss charges for spinning black hole.

\subsection{The constant time hypersurfaces}
\tsgrn
The results presented in this section holds for Schwarzschild, Schwarzschild--de Sitter and Schwarzschild--Anti de Sitter spacetimes.
\cbk Consider the metric with the standard variables
$(t,r,\theta,\phi)$, whose linear element is given by the formula:
\begin{equation}
\label{schwarzschilddesitter}
\mathrm{d} s^2 = -f(r)\mathrm{d} t^2+ \frac{\mathrm{d} r^2}{f(r)}+r^2\mathrm{d} \Omega^2\,,
\end{equation}
where $f(r)=1-\frac{2m}{r}-\bmac r^2$, and we assume that $f(r)>0\,.$ 
Parameter $\bmac $ is a scaled cosmological constant $\bmac =\frac{\Lambda}{3}$, where $s=\pm1$. The foliation with spatial hypersurfaces $\Sigma_t:t=\textrm{const}$ is examined. Three-dimensional Riemannian metric induced on $\Sigma_t$  has the following form:
\begin{equation}
\label{eq:3dSchw}
\mathrm{d} s_3^2=\frac{\mathrm{d} r^2}{f(r)}+r^2\mathrm{d} \Omega\,.
\end{equation}
The Cotton tensor for the metric
(\ref{eq:3dSchw}) is equal to zero, which (in three dimensions) is a necessary and sufficient condition for conformal flatness. Using one-dimensional coordinate transformation
\begin{equation}
\log R = \int\frac{\mathrm{d} r}{r \sqrt{1-\frac{2m}{r}-\bmac r^2}}\,, \label{eq:schwarzschild_conformal_transform}
\end{equation}
we can transform (\ref{eq:3dSchw}) into conformally flat form
\begin{equation}
\mathrm{d} s_3^2 = \frac{r^2(\log R)}{R^2}\left[ \mathrm{d} R^2 + R^2\mathrm{d} \Omega \right]\,.
\end{equation}
$I(E,\mathcal{S})$ is our conserved quantity which has an interpretation as a mass. $\mathcal{S}$ is a CKV associated with conformal rescaling. In $(R, \theta, \phi)$ coordinates in which the metric is conformally flat, $\mathcal{S}$ has the form $\mathcal{S}=R \partial_{R}$. Transforming into $(r, \theta, \phi)$ coordinates, we have
\begin{equation}
\mathcal{S}= r\sqrt{f(r)} \partial_{r}\, . \label{eq:S_in_schwarz}
\end{equation}
The chosen data is time-symmetric. It means that the second fundamental form is identically equal to zero:
\begin{equation}
K_{ij}=0 \, .
\end{equation}
The theorems \ref{theoremEBlambda} and \ref{divergenceTheorem} give
\begin{eqnarray}
E_{ij}&=&-\overset{3}{R}{}_{i j} +\frac{2}{3}\Lambda \label{eq:E_schwarz}
\gamma_{ij} \, , \\
B_{ij}&=&0 \, ,\\
{E^{ij}}_{|i}&=&0 \, . \label{eq:div_E_schwarzschild}
\end{eqnarray}
According to equation (\ref{divergencecalculation}) and the comment below, the result (\ref{eq:div_E_schwarzschild}) guarantees that the quasi-local mass does not depend on the choice of integration surface\footnote{The two-dimensional surface has to be homotopic to a round sphere -- an orbit of rotational symmetry of the Schwarzschild--de Sitter solution. In particular, it is the same for all $r=\mathrm{const}$ surfaces.}. The non-zero components of electric part of Weyl tensor do not depend on cosmological constant
\begin{equation}
E ^{r}{}_{r} = \frac{2m}{r^3}\,, \label{eq:E_r_r_schwarz}
\end{equation}
\begin{equation}
E ^{\theta}{}_{\theta} = -\frac{m}{r^3}\,,
\end{equation}
\begin{equation}
E ^{\phi}{}_{\phi} = -\frac{m}{r^3}\,.
\end{equation}
The equations (\ref{eq:S_in_schwarz}) and (\ref{eq:E_r_r_schwarz}) enable one to calculate the quasi local mass $\mathcal{M}:=\frac{1}{8 \pi} I(E,\mathcal{S})$:
\begin{equation}
\begin{aligned}
\mathcal{M} = \frac{1}{8 \pi} \int_{S(r)} E ^{r}{}_{r} X^r \sqrt{\gamma} \mathrm{d} \theta \mathrm{d} \phi =\frac{1}{8 \pi} \int_{S(r)}
\frac{2m}{r^3}r\sqrt{f(r)} \frac{1}{\sqrt{f(r)}}r^2\sin\theta \mathrm{d} \theta
\mathrm{d} \phi= m\,.
\end{aligned}
\end{equation}
The obtained result is comparable with the ADM type quasi-local mass. We will use formulas from
\cite{3} to calculate ADM type quasi-local mass relative to any vector field $X$
and any reference space. Let $\gamma _{ij}$ denote the metric at
$\Sigma_s$, the reference spacetime $\beta_{ij}$ is the de Sitter spacetime.  Assuming the above, the ADM type quasi-local mass is expressed by the formula:
\begin{equation}
M _{\textrm{ADM}} = \frac{1}{16\pi} \int_{S(r)} \left( \mathbb{U}^r +
\mathbb{V}^r \right) \,,
\end{equation}
where:
\begin{equation}
\mathbb{U}^i(V) = 2 \sqrt{\textrm{det}\gamma} \left[ V
\gamma^{i[k}\gamma^{j]l}\bar{D}{}_{j}\gamma_{kl}
+ D ^{[i}V \gamma ^{j]k}e _{jk}\right]\,,
\end{equation}
\begin{equation}
\mathbb{V}^l(Y) = 2 \sqrt{\textrm{det}\gamma} \left[ (P ^{l}{}_{k}-\bar{P} {}^{l}{}_{k})Y^k -
\frac{1}{2}Y^l \bar{P}{} ^{mn}e _{mn}+\frac{1}{2}Y^k \bar{P}{}
^{l}{}_{k}\beta^{mn}e _{mn}\right]\,,
\end{equation}
\begin{equation}
e _{ij}:=\gamma _{ij}-\beta _{ij}\,,
\end{equation}
\begin{equation}
P ^{lk} := \gamma ^{lk}\textrm{tr}_\gamma K - K
^{lk}\,,\quad\quad\textrm{tr}_\gamma K:=\gamma ^{lk}K
_{lk}\,,
\end{equation}
similarly for objects defined in the background space:
\begin{equation}
\bar{P} {}
^{lk}:=\beta ^{lk}\textrm{tr}_\beta \bar{K} - \bar{K}{}
^{lk}\,,\quad\quad\textrm{tr}_\beta\bar{K}:=
\beta^{lk}\bar{K}{} _{lk}\,.
\end{equation}
Form of the Hamiltonian field $X$:
\begin{equation}
X = V n^\mu \partial_\mu + Y^k\partial_k = \frac{V}{N}\partial_0 + \left(Y^k -
\frac{V}{N}N^k\right) \partial_k	\,.
\end{equation}
For mass (energy), we use $X=\partial_0$, so $V=N$ and $Y^k =
N^k$. The metric $\gamma$ is given by the equation (\ref{eq:3dSchw}). The reference frame is simply given by 
\begin{equation}
\beta _{ij} = \textrm{diag} \left(\frac{1}{1-\bmac r^2},r^2,r^2\sin^2\theta
\right)\,,
\end{equation}
hence:
\begin{equation}
e _{ij} = \textrm{diag} \left( \frac{1}{1-\frac{2m}{r}-\bmac r^2}-
\frac{1}{1-\bmac r^2},0,0 \right)\,.
\end{equation}
The extrinsic curvature is zero, therefore $P ^{k}{}_{l}=\bar{P} {} ^{k}{}_{l}=0
\Rightarrow \mathbb{V}^l(Y)=0\,.$\\
The expression for ADM type quasi-local mass is reduced to:
\begin{equation}
M_\textrm{ADM}=\frac{1}{8\pi}\int_{S(r)}\sqrt{\textrm{det}\gamma} \left[ N
\gamma ^{r[k}\gamma ^{j]l}\bar{D}{}_{j}\gamma _{kl}
+ D ^{[r}N\gamma ^{j]k}e _{jk}\right]\,.
\end{equation}
Which finally gives
\begin{equation}
M_\textrm{ADM} = m \,.
\end{equation}
Thus, the ADM type quasi-local mass and the "electromagnetic" quasi-local mass M are equal and they do not depend on the cosmological constant $\Lambda=\frac{3 s}{l^2}$.


\subsection{Flat hypersurfaces}
As in the previous section, we start with the metric:
\begin{equation}
\mathrm{d} s^2 = -f(r)\mathrm{d} t^2+ \frac{\mathrm{d} r^2}{f(r)}+r^2\mathrm{d} \Omega^2\,,
\end{equation}
for $f=1-\frac{2m}{r}-\bmac r^2$, $f>0$, $m \geq 0$.\\
We will examine the foliation with the hypersurfaces $\Sigma_t$, on
which induced three-dimensional metric is flat. Using the following time transformation:
\begin{equation}
t_p := t - \int \frac{\sqrt{1-f(r)}}{f(r)} \rd r\,, 
\end{equation}
we obtain the metric in the Painleve--Gullstrand form:
\begin{equation}
\label{painlevemetric}
\mathrm{d} s^2	= -f(r)\mathrm{d} t_p^2 -2\sqrt{1-f(r)}\mathrm{d} r\mathrm{d} t_p + \mathrm{d} r^2 + r^2 \Omega^2\,.
\end{equation}
The lapse $N$ and the shift vector $N^i$:
\begin{equation}
N=1\,,
\end{equation}
\begin{equation}
N_r = -\sqrt{\frac{2m}{r}+\bmac r^2}\,, \quad\quad N_\theta = 0\,, \quad\quad
N_\phi =0\, ,
\end{equation}
enables one to calculate the extrinsic curvature from the formula
\begin{equation}
K _{ij}= \frac{1}{2N} \left( N _{i|j} + N _{j|i} - \partial_{t}g _{ij}
\right)\,.
\end{equation}
Non-zero components of the extrinsic curvature are:
\begin{equation}
K _{rr} = \frac{m-\bmac r^3}{\sqrt{2mr^3+\bmac r^6}}\,,
\end{equation}
\begin{equation}
K _{\theta \theta} = -\sqrt{2mr+\bmac r^4}\,,
\end{equation}
\begin{equation}
K _{\phi \phi} = -\sqrt{2mr+\bmac r^4}\sin^2\theta\,.
\end{equation}
The trace of extrinsic curvature is equal to:
\begin{equation}
K = K _{ij}\gamma ^{ij} = - \frac{3(m+\bmac r^3)}{\sqrt{2mr^3+\bmac r^6}}\,.
\end{equation}
The electrical part of Weyl tensor, (\ref{lematElambda}), reduces to
\begin{equation}
E _{i j} = - K K _{i j} +
K _{i k}K ^{k}{}_{j}+\frac{2}{3}\Lambda \gamma_{ij}
=- K K _{i j} + K _{i k}K ^{k}{}_{j}+2\bmac \gamma_{ij}\,.
\end{equation}
We obtain the following non-zero components of the electrical part of Weyl tensor:
\begin{equation}
E _{rr} = \frac{2m}{r^3}\,,
\end{equation}
\begin{equation}
E _{\theta \theta}= - \frac{m}{r}\,,
\end{equation}
\begin{equation}
E _{\phi \phi} = - \frac{m}{r}\sin^2\theta\,.
\end{equation}
The magnetic part is zero.\\
Because the metric induced on $\Sigma_p$ is flat, we have a full set of ten conformal Killing vectors. Note that vanishing magnetic part causes that the three-dimensional covariant divergence of the electrical part vanishes (theorem \ref{divergenceTheorem}), and therefore the ``electric'' charges do not depend on the two-dimensional integration surface. The charge responsible for the mass:
\begin{equation}
\label{massESpainleve}
\mathcal{M}=\frac{1}{8 \pi} \int_{S(r)} E ^{i}{}_{j}\mathcal{S}^j \mathrm{d} S_i = \frac{1}{8 \pi} \int_{S(r)} E ^{r}{}_{r}
r^3\sin\theta \mathrm{d} \theta \mathrm{d} \phi= m\,.
\end{equation}
\noindent We compare the obtained result with the ADM type quasi-local mass. As in the previous subsection, we define:
\begin{equation}
e _{ij}:=\gamma _{ij}-\beta _{ij}\,,
\end{equation}
\begin{equation}
\mathbb{U}^i(V) = 2 \sqrt{\textrm{det}\gamma} \left[ V \gamma ^{i[k}\gamma
^{j]l}\bar{D}{}_{j}\gamma _{kl}
+ D ^{[i}V\gamma ^{j]k}e _{jk}\right]\,,
\end{equation}
\begin{equation}
\mathbb{V}^l(Y) = 2 \sqrt{\textrm{det}\gamma} \left[ (P ^{l}{}_{k}-\bar{P} {}^{l}{}_{k})Y^k -
\frac{1}{2}Y^l \bar{P}{} ^{mn}e _{mn}+\frac{1}{2}Y^k \bar{P}{}
^{l}{}_{k}\beta^{mn}e _{mn}\right]\,,
\end{equation}
\begin{equation}
X = V n^\mu \partial_\mu + Y^k\partial_k = \frac{V}{N}\partial_0 + \left(Y^k -
\frac{V}{N}N^k\right) \partial_k\,.
\end{equation}
We obtain the mass, therefore $X=\partial_0 \Rightarrow V=N$ and $Y^k=N^k$. $\mathbb{U}^l$ is zero, because in the case of foliations by flat hypersurfaces $\gamma _{ij} = \beta _{ij}$, where $e _{ij}=0$, and in the first part the covariant derivative $\bar{D}$ can be converted to $D$.\\
The non-zero components of the canonical ADM momentum $(P ^{i}{}_{j} = \delta ^{i}{}_{j}K -
K ^{i}{}_{j})$ are:
\begin{equation}
\label{ADMmomfirst}
P ^{r}{}_{r} = - \frac{2\sqrt{2m+\bmac r^3}}{r^{3/2}}\,,
\end{equation}
\begin{equation}
P ^{\theta}{}_{\theta} = - \frac{m+2\bmac r^3}{r^{3/2}\sqrt{2m+\bmac r^3}}\,,
\end{equation}
\begin{equation}
\label{ADMmomlast}
P ^{\phi}{}_{\phi} = - \frac{m+2\bmac r^3}{r^{3/2}\sqrt{2m+\bmac r^3}}\,.
\end{equation}
The background ADM momentum $\bar{P} ^{i}{}_{j}$ can be obtained from the
equations (\ref{ADMmomfirst})--(\ref{ADMmomlast}) by setting  $m=0$, hence we get:
\begin{equation}
\bar{P} ^{r}{}_{r} = \bar{P} ^{\theta}{}_{\theta}= \bar{P}
^{\phi}{}_{\phi}=-2\sqrt{\bmac }\,.
\end{equation}
The only non-omitting element in the expression for ADM type quasi-local mass is:
\begin{equation}
M_{\textrm{ADM}} = \frac{1}{16\pi} \int_{S(r)} 2\sqrt{\gamma}(P
^{r}{}_{r}-\bar{P}{}^{r}{}_{r})N^r \mathrm{d} \theta \mathrm{d} \phi\,.
\end{equation}
Explicitly,
\begin{equation}
M_{\textrm{ADM}} =  2m+\bmac r^3-\sqrt{\bmac r^3(2m+\bmac r^3)} \,.
\end{equation}
Note that for Schwarzschild spacetime ($\bmac =0$) we get $M
_{\textrm{ADM}} = 2m$, which is twice as much as the mass parameter in the metric. This is caused by too slow fall off in $r$ of the ADM momentum. In the case of foliation with surfaces of flat internal geometry (Painleve--Gullstrand foliation), the tensor of the ADM momentum behaves like $r^{-3/2}$. The usually assumed assumption for ADM momentum (see classical results in section \ref{sec:chr_class_results}) for ADM mass is the fall off like $r^{-3/2-\varepsilon}$, where $\varepsilon$
is strictly positive. The result (\ref{massESpainleve}) shows that the ``electromagnetic'' mass (at least in this case) has better properties,
because after dividing by the normalizing factor $8\pi$ accurately reproduces the parameter $m$ occurring in the metric. It is independent of the radius $r$ and the
scaled cosmological constant $\bmac $.

\subsection{Bowen--York initial data type \label{sec:Dain_procedure_to_B_Y}}
In the section we discuss an initial data which is originally done by Bowen and York \cite{bowen1980time} with the help of conformal methods. However, we use an approach to the data which is given in  \cite{dain2002new}.\\
Physically relevant initial data $(\gamma_{ab},K_{ab})$ is related with conformal data by the relations
\begin{eqnarray}
\gamma_{ij} &=& 	\varphi^4  \widetilde{h}_{ij} \, , \\
K^{ij} &=& \varphi^{-10}\widetilde{K}^{ij} \, .
\end{eqnarray}
$\widetilde{K}_{ij}$ is a symmetric, trace free tensor which fulfills
\begin{equation}
\label{eq:1}
D_i \widetilde{K}^{ij}=0 \, ,
\end{equation}
where $D_a$ is the covariant derivative with respect to $ \widetilde{h}_{ij}$. The following analog of the Hamiltonian constraint for the conformal factor $\varphi$ exist
\begin{equation}
\label{eq:conformal_factor_general}
L_{ \widetilde{h}} \varphi= -\frac{\widetilde{K}^{ij}\widetilde{K}_{ij}}{8\varphi^7},
\end{equation}
where $L_{ h}= D^i  D_i-  R/8$, $ R$ is the Ricci scalar of the metric $\widetilde{h}_{ij}$ and the indexes are moved with  $ \widetilde{h}_{ij}$. If the conditions (\ref{eq:1}) and (\ref{eq:conformal_factor_general}) holds then the physically relevant initial data $(\gamma_{ij},K_{ij})$ satisfy vacuum constraint equations without cosmological constant.\\
Remarkable simplifications on (\ref{eq:1}) and (\ref{eq:conformal_factor_general}) occur when $ \widetilde{h}_{ij}$ has a Killing vector $\eta^a$. We will assume that $\eta^a$  is hypersurface orthogonal\footnote{i.e.; it satisfies $D_i
	\eta_j=-\eta_{[i}D_{j]} \ln \eta$} and we define $\eta$   by $\eta=\eta^a
\eta^b  \widetilde{h}_{ij}$. We analyze first the momentum constraint (\ref{eq:1}). Consider the following vector field 
\begin{equation}
\label{eq:axialve}
S^i=\frac{1}{\eta} \epsilon^{ijk} \eta_j  D_k \omega, \quad
\pounds_\eta \omega =0, 
\end{equation}
where  $\pounds_\eta$ is the Lie derivative with respect
$\eta^a$ and $\epsilon_{ijk}$ is the volume element of $\widetilde{h}_{ij}$. The scalar function $\omega$ is arbitrary. In particular it does not depend on the metric $\widetilde{h}_{ij}$. It follows  that $S^a$ satisfies
\begin{equation}
\label{eq:J}
\pounds_\eta S^i=0, \quad S^j\eta_j=0, \quad D_k S^k=0.
\end{equation}
Using  the Killing equation $D_{(i}\eta_{j)}=0$,  the fact that
$\eta^a$ is hypersurface orthogonal and equations (\ref{eq:J}) we
conclude that the  tensor 
\begin{equation}
\label{eq:axialpsi}
\widetilde{K}^{ij}=\frac{2}{\eta} S^{(i} \eta^{j)},
\end{equation}
is trace free and satisfies (\ref{eq:1}).
The square of $\widetilde{K}^{ij}$ can be written in terms of $\omega$
\begin{equation}
\label{eq:K2_in_omega}
\widetilde{K}^{ij}\widetilde{K}_{ij}=2\frac{D_k\omega D^k\omega}{\eta^2}.
\end{equation}
\subsection{Example: Bowen--York spinning black hole}
The example originally has been given in \cite{bowen1980time}. It is a Bowen--York type initial date with a Killing symmetry. The data is characterized by a flat conformal metric
\begin{equation}
\label{eq:flat_metric}
\widetilde{h} = \mathrm{d} R^2+R^2 \mathrm{d} \theta^2+R^2\sin^2\theta \mathrm{d} \phi^2.
\end{equation}
with the following generating function $\omega$
\begin{equation}
\label{eq:omega_B_Y}
\omega_{BY}=J(\cos^3\theta-3\cos\theta).
\end{equation}
The Killing vector used in data construction is
\begin{equation}
\eta=\frac{\partial}{\partial \phi}
\end{equation}

Using the procedure described in section \ref{sec:Dain_procedure_to_B_Y} we obtain the only non-vanishing component of $\widetilde{K}_{ij}$ is the following
\begin{equation}
\widetilde{K}_{R \phi}=\frac{3 J \sin (\theta)^{2}}{R^{2}}
\end{equation} 
Equations (\ref{eq:conformal_factor_general}) and (\ref{eq:K2_in_omega}) enables one to obtain the relation for the conformal factor for the  flat metric (\ref{eq:flat_metric}) in terms of the derivatives of $\omega$
\begin{equation}
\label{eq:conformal_factor_flat}
\Delta \varphi =-\frac{(r\partial_r\omega)^2 +
	(\partial_\theta \omega)^2}{4 R^6 \sin^4\theta \varphi^7}
\end{equation}
where $\Delta$ is the flat Laplacian in the spherical coordinates
$(R, \theta, \phi)$. The following boundary condition are used
\begin{equation}
\label{eq:boundary_conditions}
\lim_{R\rightarrow \infty} \varphi =1, \quad  \lim_{R\rightarrow
	0} R\varphi=\frac{M}{2}, 
\end{equation}
where $M$ is a positive constant called bare mass. We have calculated an asymptotic behavior of the solution for large $R$. The first few terms of asymptotic series of the conformal factor are the following
\begin{equation}
\varphi_{BY}=1+\frac{M}{2 R}+\frac{p(\theta)}{R^{4}}+O\left(\frac{1}{R^5}\right) \label{eq:conformal_factor_BY}
\end{equation}
where
\begin{equation}
p(\theta)=\frac{J^{2}}{8}\left(3 \cos (\theta)^{2}-1\right)-\frac{J^{2}}{8} \, .
\end{equation}
The charges (\ref{eq:def_Q_E}) and  (\ref{eq:def_Q_B}) requires gravitoelectromagnetic tensors for the solution. The  gravitomagnetic tensor is relatively simple
\begin{eqnarray}
B_{RR}&=&-\frac{6 J\left(\varphi_{BY} \cos \theta -2 \sin \theta \partial_{\theta} \varphi_{BY} \right)}{\varphi_{BY}^{5} R^{4}} \\
B_{R \theta}&=&-\frac{3\left(2 R \partial_{R}\varphi_{BY}+\varphi_{BY}\right) J \sin \theta}{\varphi_{BY}^{5} R^{3}} \\
B_{\theta \theta}&=&\frac{3 J \cos \theta}{\varphi_{BY}^{4} R^{2}} \\
B_{\phi \phi}&=&\frac{3 J \sin^2 \theta \left(\varphi_{BY} \cos \theta -4\sin \theta \partial_{\theta}\varphi_{BY} \right)}{\varphi_{BY}^{5} R^{2}}
\end{eqnarray}
The non-vanishing components of electric part of Weyl tensor are 
\begin{eqnarray}
E{{}^{R}}_{R}&=&\frac{2 M}{R^{3}\left(1+\frac{M}{2 R}\right)^{6}}+\frac{1}{2} \frac{21 J^{2} \cos^2 \theta -11 J^{2}}{R^{6}}+O\left(\frac{1}{R^7}\right)\\
E{{}^{R}}_{\theta}&=&O\left(\frac{1}{R^7}\right)\\
E{{}^{\theta}}_{\theta}&=&\frac{M}{R^{3}\left(1+\frac{M}{2 R}\right)^{6}}-\frac{1}{2} \frac{\left(3 \cos^2 \theta+2\right) J^{2}}{R^{6}}+O\left(\frac{1}{R^7}\right)\\
E{{}^{\phi}}_{\phi}&=&\frac{M}{R^{3}\left(1+\frac{M}{2 R}\right)^{6}}-\frac{1}{2} \frac{\left(18 \cos^2 \theta-13\right) J^{2}}{R^{6}}+O\left(\frac{1}{R^7}\right)
\end{eqnarray} 
Angular momentum $I(B,\mathcal{K}_{z})$ is generated by conformal acceleration in $z$ axis
\begin{equation}
\mathcal{K}_z =
\frac{1}{2}R^2\cos\theta\partial_R+\frac{1}{2}R\sin\theta\partial_\theta\,.
\end{equation}
The angular momentum is finally given by
\begin{eqnarray}
-\frac{1}{8 \pi} I(E,\mathcal{K}_{z}) &=&-\frac{1}{8 \pi}  \int_{S(R)}\left(B ^{R}{}_{R} \mathcal{K}_z ^R+B ^{R}{}_{\theta} \mathcal{K}_z ^\theta \right) \varphi_{BY}^6 R^2 \sin \theta \mathrm{d} \theta \mathrm{d} \phi \nonumber \\
&=& J +\frac{ M J}{2 R}+\frac{4}{5} \frac{J^{3}}{R^{4}}+O\left(\frac{1}{R^5}\right)
\end{eqnarray}
The quasi-local mass\footnote{We recall the corresponding to the mass conformal Killing vector field is simply $X=R \partial_{R}$. }  is calculated for $R=\mathrm{const}$ surfaces. It reads
\begin{eqnarray}
\mathcal{M} &=& \frac{1}{8 \pi} \int_{S(R)} E ^{R}{}_{R} X^R \sqrt{\gamma} \mathrm{d} \theta \mathrm{d} \phi \nonumber \\
&=& M-\frac{ J^{2}}{R^{3}}+O\left(\frac{1}{R^4}\right)
\end{eqnarray}
\section{Asymptotic charges}
\label{sec:asymptotic_charges}
Brief review of classical and recent results for asymptotic charges is given in Appendix \ref{sec:chr_class_results}.
\subsection{Existence of asymptotic $I(E,X)$ and $I(B,X)$  }
In the section \ref{sec:Instant_charges} the definition of instantaneous charges has been given. The conservation laws for instantaneous charges have been provided in proposition \ref{prop:IC_conservation}. We generalize the concepts by considering objects which fulfills the assumptions in the proposition only in asymptotic regime.  For convenience, we define a symmetric tensor which describes deviation of a vector field $Y$ from being CKV
\begin{equation} \label{eq:def_V}
V _{ij}(Y) := Y _{(i|j)} - \frac{1}{3} Y ^{k}{}_{|k} \gamma _{ij}\,.
\end{equation}
If $X$ is a conformal Killing vector for $\gamma$ then $V _{ij}(X,\gamma)=0$. With the help of $V$, the definition of asymptotic conformal Killing vector field reads
\begin{defi}(asymptotic CKV) \label{def:def_aCKV}
Vector field X will be called asymptotic conformal Killing vector field iff
\begin{equation*}
\lim_{r \to \infty} V_{ij}(X)=0 \, ,
\end{equation*}
 for all $i,j$. The limit $\lim_{r \to \infty}$ is understood as the spatial infinity regime.
\end{defi}
It enables one to define asymptotic instantaneous charges:
\begin{defi}[asymptotic instantaneous charges]
	Consider a given spatial hypersurface $\Sigma$ equipped with an asymptotically flat\footnote{We assume there exist a coordinate chart that the metric tends to the Euclidean metric at spatial infinity.}, Riemannian metric $\gamma$. Let $A(r) \subset \Sigma$ be an one parameter family of a two-dimensional closed surfaces such that $\lim\limits_{r \to \infty} A(r)$ represents a topological sphere at spatial infinity. We can define the following asymptotic instantaneous charges:
	\begin{eqnarray}
	I_{as}(E,X) &:=& \lim_{r \to \infty} \int_{S(r)} E ^{i}{}_{j}X^j dS_i\,, \label{eq:def_asymp_I_E}\\
	I_{as}(B,X) &:=& \lim_{r \to \infty} \int_{S(r)} B ^{i}{}_{j}X^j dS_i\,, \label{eq:def_asymp_I_B}
	\end{eqnarray}
	where $X^{j}$ is an asymptotic conformal Killing vector field (see definition \ref{def:def_aCKV}). 
\end{defi} 
 Let us recall the reasoning which leads to conservation laws for instantaneous charges. See the equation (\ref{powierzchnieS}) and comments nearby. Analogically, the boundary integrals, (\ref{eq:def_asymp_I_E}) and (\ref{eq:def_asymp_I_B}), would be well-defined if the divergences, $(\sqrt{\gamma}E ^{ij}X _{j}) _{,i}$ and $(\sqrt{\gamma}B ^{ij}X _{j})_{,i}$ respectively, will be integrable. We have 

\begin{equation}
\label{eq:asympt_div_E}
\begin{aligned}
(\sqrt{\gamma}E ^{ij}X _{j}) _{,i}
&=\sqrt{\gamma} E^{ij}(X _{(i|j)} -\frac{1}{3}X ^{k}{}_{|k}\gamma _{ij} )+ \sqrt{\gamma} E ^{i}{}_{j|i}X^j \\
&=\sqrt{\gamma} E^{ij} V _{ij}(X)+\sqrt{\gamma}X^j(K\wedge B) _{j}
\end{aligned}
\end{equation}
where (\ref{eq:div_E_KwB}) is used. Analogical calculations for $(\sqrt{\gamma}B ^{ij}X _{j})_{,i}$  gives 
\begin{equation}
\label{eq:asympt_div_B}
(\sqrt{\gamma}B ^{ij}X _{j}) _{,i} =\sqrt{\gamma} B^{ij} V _{ij}(X)-\sqrt{\gamma}X^j(K\wedge E) _{j}
\end{equation}
The asymptotic instantaneous charges will be finite if the appropriate divergences are integrable at infinity, i.e.
\begin{eqnarray}
\label{asymptotycznywarunekE}
 E^{ij} V _{ij}(X)+X^j(K\wedge B) _{j}&=&O(r^{-3-\varepsilon})\,,\\
\label{asymptotycznywarunekB}
 B^{ij} V _{ij}(X)-X^j(K\wedge E) _{j}&=&O(r^{-3-\varepsilon})\,,
\end{eqnarray}
where $\varepsilon>0$. \\
Note that the basic conformal Killing vectors have different asymptotic relative to $r$,
using (\ref{CKVT})--(\ref{CKVS}) we have:
\begin{itemize}
	\item $\mathcal{T}[k] = O(1)\,,$
	\item $\mathcal{R}[k] = O(r)\,,$
	\item $\mathcal{S} = O(r)\,,$
	\item $\mathcal{K}[k] = O(r^2)\,.$
\end{itemize}
Generators of proper conformal transformations behave like $r^2$,
therefore charges defined with $\mathcal{K}_k$ (e.g. angular momentum)
impose the strongest conditions on the asymptotic in the equations
(\ref{asymptotycznywarunekE})--(\ref{asymptotycznywarunekB}). For convenience, we can rewrite the asymptotic of CKV in more compact form
\begin{equation}
	X=O(r^{q})\, , \quad q \in \{0,1,2\} \label{eq:compact_CKV}
\end{equation}
The conditions (\ref{asymptotycznywarunekE}) and (\ref{asymptotycznywarunekB}) can be interpreted in various ways.\\
For example, if we assume that $X$ is a CKV ($V _{ij}(X)=0$) and the metric is asymptotically flat i.e. the conformal factor of the metric tends to unity in spatial infinity then the conditions (\ref{asymptotycznywarunekE}) and (\ref{asymptotycznywarunekB}) are equivalent to
\begin{eqnarray}
(K \wedge E) _{j}&=&O(r^{-3-q-\varepsilon})\,,\\
(K \wedge B) _{j}&=&O(r^{-3-q-\varepsilon})\,.
\end{eqnarray} 
where $q$ depend on the type of charge\footnote{We remind the physical interpretation of the charges is given in section \ref{sec:physical_interpretation}.} and associated CKV (\ref{eq:compact_CKV}).    

We can also consider surfaces $\Sigma$, whose metric is only asymptotically conformally flat.
\begin{theorem}
	\label{asymptotic}
	Let the metric on the hypersurface $\Sigma$ be in the form\footnote{The metric of such a form corresponds to slightly disturbed Schwarzschild spacetime
		(see Schwarzschild metric expressed in isotropic variables, see
		(\ref{eq:schwarzschild_conformal_transform})).}:
	\begin{equation}
	\gamma _{ij} = \left( 1+ \frac{M}{2r} \right)^4 \left( \eta _{ij} + h
	_{ij} \right)\,,
	\end{equation}
	where $\eta _{ij}$ is a three-dimensional Euclidean metric and \tsgrn $M$ is a bare mass. \cbk 
	We assume the following behavior of presented objects at $r \rightarrow \infty\;:$
	\begin{equation}
	\begin{aligned}
	h _{ij} =O\left( r^{d}\right)\,, \quad (\Gamma _{\eta+h}) ^{a}{}_{ij} = O\left( r^{c} \right)\,, \quad K _{ij} =O\left( r^{k}\right)\,,\\
	E _{ij} =O\left( r^{e}\right)\,, \quad B _{ij} =O\left( r^{b}\right)\,, \quad X=O(r^{q})\, , \quad q \in \{0,1,2\} \, , 
	\end{aligned}
	\end{equation}
	where $c<d<1$ and X is a CKV for $\eta$ defined by (\ref{eq:compact_CKV}) and the comments nearby. \\
	If 
	\begin{equation} \label{eq:asympt_conditions_aICe}
	\begin{aligned}
	e+c+q<-3 \, , \\
	e+q+d<-3 \, ,\\
	c+k+b<-3 \, ,
	\end{aligned}
	\end{equation}
	then the asymptotic charge
	\begin{equation} \label{eq:aICe_thm}
	I_{as}(E,X) = \lim_{r \to \infty} \int_{S(r)} E ^{i}{}_{j}X^j dS_i\,,
	\end{equation}
	is integrable. Analogically, the following conditions
		\begin{equation}
	\begin{aligned}
	b+c+q<-3 \, , \\
	b+q+d<-3 \, ,\\
	c+k+e<-3 \, ,
	\end{aligned}
	\end{equation}
	guarantees that the asymptotic instantaneous charge
	\begin{equation}\label{eq:aICb_thm}
	I_{as}(B,X) = \lim_{r \to \infty} \int_{S(r)} B ^{i}{}_{j}X^j dS_i\,,
	\end{equation}
	is finite.
\end{theorem}
\noindent\textit{Proof:}\\
The aim of the proof is to show that at $r \rightarrow \infty$
divergences given by equations (\ref{eq:asympt_div_E}), (\ref{eq:asympt_div_B}) are integrable at infinity. It leads to asymptotic given by (\ref{asymptotycznywarunekE}) and (\ref{asymptotycznywarunekB}). For the metric $\eta$, (\ref{CKVT})--(\ref{CKVS}) are exact solutions of CKV equation ($V_{ij}=0$). Asymptotic of such CKV is given by (\ref{eq:compact_CKV}). Let us examine expressions containing tensor $V _{ij}$. The tensor $V$ behaves in conformal transformation as follows:
\begin{equation}
V _{ij}\left(\Omega^4 g, X\right) = \Omega^4 V _{ij}(g, X)\,,
\end{equation}
where $g$ is any metric. In the case, $\Omega = 1+\frac{M}{2r}$ the leading therm of conformal factor is $O(1)$. We assume that the field $X$ is a conformal Killing vector for a flat, three-dimensional metric $\eta$, thus the following equation is satisfied:
\begin{equation} \label{eq:CKV_assump_proof}
X _{(i,j)}-\frac{1}{3}X ^{k}{}_{,k}\eta _{ij}=0\,.
\end{equation}
Hence,
\begin{equation}
V(\eta+h,X) = -\Gamma ^{m}{}_{ij}X_m-\frac{1}{3}X ^{k}{}_{,k}h
_{ij}-\frac{1}{3}\Gamma ^{k}{}_{mk}X^m(\eta _{ij}+h _{ij})\,.
\end{equation}
The Christoffel symbols appearing in the above formula come from the metric $\eta
+h$, we get
\begin{equation}
V \left( \Omega^4(\eta+h),X \right) = \Omega^4 V (\eta+h,X) =
O(r^{v})\,.
\end{equation}
where
\begin{equation}
v = \max \left(c+q,q+h-1,q+h \right)= \max \left(c+q,q+h \right)
\end{equation}
 It leads to
\begin{eqnarray} \label{eq:asymp_E_V}
 E
^{ij}V _{ij}(g,X) &=&O(r^{e})\cdot O(r^{v}) = O(r
^{e+v})\,,\\
 \label{eq:asymp_B_V}
 B
^{ij}V _{ij}(g,X) &=& O(r^{b})\cdot O(r^{v}) = O(r
^{b+v})\,.
\end{eqnarray}
Next, we check the asymptotic of the terms containing the extrinsic curvature:
\begin{eqnarray} \label{eq:asymp_K_B}
X^j(K \wedge B)_{j} &=& O(r^c) \cdot O(r^{k}) \cdot
O(r^{b})=O(r^{c+k+b})\,,\\
 \label{eq:asymp_K_E}
X^j(K \wedge E)_{j} &=& O(r^c) \cdot O(r^{k}) \cdot
O(r^{e})=O(r^{c+k+e})\,.
\end{eqnarray}
Comparing (\ref{asymptotycznywarunekE}),(\ref{eq:asymp_E_V}) and (\ref{eq:asymp_K_B}), asymptotic charge
\begin{equation}
I_{as}(E,X) = \lim_{r \to \infty} \int_{S(r)} E ^{i}{}_{j}X^j dS_i\,,
\end{equation}
will be integrable if
\begin{eqnarray}
\max(e+c+q,e+q+h,c+k+b)&<&-3
\end{eqnarray}
With the help of (\ref{asymptotycznywarunekB}),(\ref{eq:asymp_B_V}) and (\ref{eq:asymp_K_E}), for the charge
\begin{equation}
I_{as}(B,X) = \lim_{r \to \infty} \int_{S(r)} B ^{i}{}_{j}X^j dS_i\,,
\end{equation}
 we can formulate analogical condition 
 \begin{eqnarray}
 \max(b+c+q,b+q+h,c+k+e)&<&-3
 \end{eqnarray}
 \qed

\paragraph{Remark:} Let us note the above proof holds even if we weak the assumption that $X$ is an exact CKV. Precisely,  the condition (\ref{eq:CKV_assump_proof}) can be substituted by 
\begin{equation} \label{eq:aCKV_r_p}
V _{ij}(X) = O(r^p),, 
\end{equation}
where $p\leq0$ will be specified below.
If we extend (\ref{eq:asympt_conditions_aICe}) by the condition
\begin{equation} \label{eq:aditional_aICe_integrability_condition}
e+p<-3
\end{equation}
 then the asymptotic charge (\ref{eq:aICe_thm}) is integrable for aCKV (\ref{eq:aCKV_r_p}). The analogical condition
 \begin{equation}
 b+p<-3
 \end{equation}
 enables one to generalize the theorem for the charge (\ref{eq:aICb_thm}).  

\section*{Conclusions}
We have provided relations between four dimensional charges (\ref{eq:def_I}), constructed with the help of conformal Yano--Killing tensor, and the instantaneous charges (definition \ref{def:instantaneous_charges}) after (3+1) decomposition. The construction presented here enables one to calculate twenty local charges in terms of initial data $(\gamma _{ij}, K _{ij})$ on a conformally flat spatial hypersurface $\Sigma$ immersed in a spacetime satisfying the Einstein vacuum equations with a cosmological constant.\\
\noindent In the future, we plan to perform the analysis of (3+1) decomposition of CYK tensor in curved spacetimes. For Minkowski spacetime, each CYK two-form is a linear combination of a wedge product of two conformal Killing co-vectors (CKV) or a Hodge dual of such product. This makes the (3+1) decomposition of CYK tensor simple. In particular, the choice of Cauchy surface $t=\mathrm{const}$ is natural.  In the case of curved spaces, the splitting of CYK form is much more complicated and the choice of Cauchy surface is much less intuitive. However, the CYK tensor decomposition should guarantee existence of Gaussian charges on a properly chosen surface. Deep understanding of the construction in the flat case is required for further analysis of curved spacetimes. This makes the paper valuable in the context of further research.\\
\noindent In addition, we have proved theorems explaining relation between linear momentum and angular momentum (defined as the contractions of the magnetic part of Weyl tensor with appropriate conformal Killing vectors) and the traditional ADM linear momentum and the ADM angular momentum. \\
The analysis performed for the Schwarzschild--de-Sitter spacetime showed that the mass defined as the contraction of the electrical part of the Weyl tensor with the scaling generator may have better properties than the ADM mass. In the proposed example of foliation with surfaces with flat internal geometry, the mass calculated with conformal Killing fields was not dependent on radius or cosmological constant (as opposed to ADM mass).\\
The thesis formulated in the last section shows that in certain specific cases we can use the proposed method to define asymptotic charges (even if the metric on spatial surfaces is not conformally flat).

\subsection*{Acknowledgments}
This work was supported in part by Narodowe Centrum Nauki (Poland) under Grant No. 2016/21/B/ST1/00940.\\
This material is based upon work supported by the Swedish Research Council under grant no. 2016-06596 while  one of the authors (JJ) was in residence at Institut Mittag-Leffler in Djursholm, Sweden during the Research Program: General Relativity, Geometry and Analysis: beyond the first 100 years after Einstein,
02 September - 13 December 2019.
\appendix
\section{Notation and conventions \label{sec:notations}}
For convenience we use index notation with Einstein summation convention. Signature of Lorentzian four-dimensional metric $g_{\mu \nu}$ is $(-,+,+,+)$. We distinguish five types of indices:
\begin{itemize}
    \item small Greek letters $\{\alpha, \beta, \gamma,...\}$, except $\theta$ and $\phi,$ represent four-dimensional coordinates of space-time,
    \item small Latin letters $\{i,j,k,...\}$, except $x$, $y$, $z$ and $r,$ run coordinates on three-dimensional spatial hypersurface $\Sigma_t$,
    \item capital Latin letters $\{A,B,C,...\}$ represents angular coordinates on two-dimensional sphere,
    \item $\{x,y,z\}$ represents a set of Cartesian coordinates,
    \item $\{r,\theta,\phi\}$ represents a set of spherical coordinates.	
\end{itemize}
The metric induced on a three-dimensional spatial hypersurface is denoted by $\gamma_{ij}$. $\eta_{\mu \nu}$ and $\eta_{AB}$ respectively mean a four-dimesional Minkowski metric and a metric on two-dimesional sphere of radius $r,$ i.e. $\eta_{AB} \mathrm{d} x^A \mathrm{d} x^B=r^2 \mathrm{d}\theta^2+r^2 \sin^2 \theta \mathrm{d} \phi^2$. $\nabla$ and $D$ mean four-dimensional and three-dimensional covariant derivatives compatible with the metrices $g_{\mu \nu}$ and $\gamma_{i j}$ respectively.
We will also use shortened, symbolic notation between indices in which semicolon ($;$) means a covariant derivative $\nabla$ for space-time, vertical line ($|$) is a covariant derivative $D$ on three-dimensional hypersurfaces, and a~double vertical line ($||$) means a covariant derivative on a two-dimensional sphere. Partial derivative $\partial$ is denoted by comma ($,$). For example, $A_{\mu \nu;\gamma}=\nabla_{\gamma} A_{\mu \nu}$. Symmetrization and antisymmetrization of indices $\alpha,\beta$ we write
respectively as $(\alpha \beta)$ and $[\alpha \beta]$, we assume that both of these
operations contain a numerical factor depending on the number of indices that
include, in particular:
\[
A _{(ij)} = \frac{1}{2}A _{ij} + \frac{1}{2}A _{ji}\,,
\]
\[
A _{[ij]} = \frac{1}{2}A _{ij} - \frac{1}{2}A _{ji}\,,
\]
\[
A _{ij} = A _{(ij)}+A _{[ij]}\,.
\]
For the antisymmetric tensor, we accept the convention:
\[
\sqrt{|g|} \varepsilon^{1 2 \dots n} = 1\,.
\]

%
\noindent In the case of calculation related to the (3 + 1) decomposition, we will write three over tensors specified on the spatial hypersurface $\Sigma_t$ and four
for objects in four-dimensional spacetime. To simplify the notation,
we omit the index in situations, where it is clearly derived from the context. The geometric layout of units was adopted throughout the work $c=G=1$.

\noindent \textbf{Index of notation} \\
$g _{\mu\nu}$ -- four-dimensional metric tensor, \\
$\eta_{\mu\nu}$ -- metrics of Minkowski spacetime,\\
$R _{\alpha \beta \mu \nu}$ -- Riemann tensor,\\
$R _{\alpha \beta}$ -- Ricci tensor, \\
$R$ -- curvature scalar, \\
$W _{\alpha \beta \mu \nu}$ -- Weyl tensor,\\
$E _{\mu \nu}$ -- electric part of Weyl tensor, \\
$B _{\mu \nu}$ -- magnetic part of Weyl tensor, \\
$\Lambda$ -- cosmological constant, \\
$T _{\mu \nu}$ -- energy-momentum tensor, \\
$S(r)$ -- two-dimensional sphere with radius $r$,\\
$\textrm{d}\Omega^2 = r^2\textrm{d}\theta^2 + r^2\sin^2\theta \textrm{d}\phi^2$.\\

The conformal Killing vectors (CKV): \\
$\mathcal{T}_k$ -- translation generators, \\
$\mathcal{R}_k$ -- rotation generators,\\
$\mathcal{K}_k$ -- generators of proper conformal transformations, \\
$\mathcal{S}$ -- scaling generator.\\

(3 + 1) Decomposition: \\
$\Sigma$ --  spatial hypersurface,\\
$N$ -- lapse function, \\
$N^k$ --  shift vector, \\
$\gamma _{ij}$ -- Riemann metric on a hypersurface $\Sigma$,\\
$K _{ij}$ -- tensor of the extrinsic curvature, \\
$K$ -- trace of the tensor of the extrinsic curvature $K=K _{ij}g ^{ij}$,\\
$P _{ij}$ -- canonical  ADM momentum $P _{ij} = K _{ij} - \gamma _{ij}K$,\\
$n^k$ -- normalized normal vector, \\
$	(A \wedge B)_a :=
\varepsilon _{a}{} ^{bc}A _{b}{}^dB _{dc}$.
\section{Conformal Killing fields}
\label{sec:CKV}
A vector field $X$ is a Conformal Killing field (CKV) if it satisfies the equation:
\begin{equation}
\mathcal{L}_X g = \lambda g \,, \label{eq:kill_lie}
\end{equation}
where $g$ is a metric tensor, and $\mathcal{L}$ is the Lie derivative.\\
If the Levi--Civita connection preserves $g$, the above equation becomes equivalent
to:
\begin{equation}
\label{killinga}
\nabla_{(\mu}X _{ \nu)}= \lambda g _{\mu \nu} \,,
\end{equation}
where $\lambda$ is an arbitrary function which is related to the conformal factor. The analogous CKV equation (\ref{eq:kill_lie}) holds also for any dimension. In general case
function $\lambda$ depends on the field $X$, and let $n$ be a dimension
of our manifold. Trace of the equation (\ref{killinga}) yields:
\begin{equation}
\label{ckillambda}
2X^\mu{}_{;\mu} = n \lambda \quad\Rightarrow\quad \lambda =
\frac{2}{n}X^\mu{}_{;\mu}\,.
\end{equation}
CKV remains invariant under the influence of conformal transformations. More precise information are contained in the following lemma:
\begin{lemat}
    \label{CKVflatlemma}
    If $X^a$ is a conformal Killing field for a metric $g _{ij}$ and
    conformal factor $\lambda$, than is also a conformal Killing field for the metric
    $\tilde{g} _{ij}= e^{2\Omega} g _{ij}$ and conformal factor $\tilde{\lambda}
    = \lambda + 2\Omega _{,a}X^a$.
\end{lemat}
If $\lambda=0$, we obtain Killing field which is an isometry generator for the conformal metric.  
The conformally flat three-dimensional Euclidean space has ten linearly independent conformal Killing fields: three fields corresponding to translation generators $\mathcal{T}_k$, three
corresponding to rotation generators $\mathcal{R}_k$, three conformal transformations $\mathcal{K}_k$ and the scaling field $\mathcal{S}$. In the Cartesian coordinate system, expressions for the fields take the following form:
\begin{eqnarray}
\label{CKVT}
\mathcal{T}[k]&=&\frac{\partial}{\partial x^k}\,,\\
\label{CKVR}
\mathcal{R}[k]&=&\varepsilon_k{}^{ij}x_i\frac{\partial}{\partial x^j}\,,\\
\label{CKVK}
\mathcal{K}[k]&=&x_k\mathcal{S}-\frac{1}{2}r^2\frac{\partial}{\partial x^k}\,,\\
\label{CKVS}
\mathcal{S}&=&x^k \frac{\partial}{\partial x^k}\,,
\end{eqnarray}
where $r^2=x^2+y^2+z^2$. Note that the above definitions in a natural way distinguish one point --- the center of the coordinate system.
\section{Brief review of classical and recent results for asymptotic charges \label{sec:chr_class_results}}

The formulation presented below follows the results obtained by Chruściel in \cite{chrusciel1986}. Let us assume that $\Sigma$ is a spacelike hypersurface extending up to infinity in an asymptotically flat space-time, where ``asymptotic flatness"
is to be understood as follows: outside a world tube there exists
a coordinate system such that
\begin{equation}
g_{\mu\nu}=\mathring{\eta}_{\mu\nu}+h_{\mu\nu},
\end{equation}
where $\mathring{\eta}_{\mu\nu}$ is the Minkowski metric, and
$h_{\mu\nu}$ satisfies
\begin{equation}
|h_{\mu\nu}| \leq  C/r^\alpha\;,\qquad |h_{\mu\nu,\sigma}| \leq
C/r^{\alpha+1}, \label{eq:chr_cond}
\end{equation}
for some $\alpha$ to be specified later.\\
We will perform asymptotic analysis in the case of a fixed Cauchy
hypersurface $\Sigma$. Before giving the precise statement of the
theorems, it is useful to introduce first some terminology.
Suppose one is given a pair $(g,\Phi)$, where
\begin{enumerate}
    \item $g$ is a Riemannian metric on a three dimensional manifold $N$,
    $N$ diffeomorphic to $\mathbb{R}^3\setminus B(R)$, where $B(R)$ is a
    closed ball ($N$ can be thought of as one of (possible many)
    ``ends" of $\Sigma$\ ).
    \item $\Phi$ is a coordinate system in the complement of a compact
    set $K$ of $N$ such that, in local coordinates $\Phi^i (p)=x^i$
    the metric takes the following form:
    \begin{equation}
    g_{ij}=\delta_{ij}+k_{ij},
    \end{equation}
    and $k_{ij}$ satisfies
    \begin{equation}
    \forall_{i,j,k,x}\ |k_{ij} (x)| \leq \ C/(r+1)^{\alpha}\ \, \ \
    |\partial k_{ij}/\partial x^k(x)| \leq \ C/(r+1)^{\alpha+1}\ \,
    \end{equation}
    for some constant $C \in \mathbb{R}$ and $r = (\sum_{i} (x^i)^2\ )^{1/2} $. Such a pair $(g,\Phi)$ will be called
    $\alpha$--admissible.
\end{enumerate}
Let us restate the remaining boundary
conditions (\ref{eq:chr_cond}) in the ADM language:
\begin{eqnarray}
&
\forall i,j,x \quad|(N-1)(x)| \le C/(r+1)^\alpha\;,\qquad
|N^i(x)|\le C/(r+1)^\alpha\;,
& \nonumber \\
&
|N_{,i}(x)| \le C/(r+1)^{\alpha+1}\;,\quad
|P_{ij}(x)|\le C/(r+1)^{\alpha+1}\;,\quad |N^i{}_{,j}(x)|\le
C/(r+1)^{\alpha+1}\;. \phantom{XX}
& \label{eq:chr_cond_ADM}
\end{eqnarray}
In \cite{chrusciel1986}, Chruściel proves the following theorems:\\
\begin{theorem}
    \label{thm:classical_ADM}
    Suppose that
    \begin{enumerate}
        \item $(g,\Phi)$ is $\alpha$--admissible, with $\alpha > 1/2$,
        \item the conditions (\ref{eq:chr_cond_ADM}) are satisfied,
        \item $(g_{ij}, P_{ij})$ satisfy the constraint equations, with
        integrable sources. 
    \end{enumerate}
    Let $S(R)$ be any one-parameter family of differentiable spheres,
    such that $r(S(R))=\min _{x\in S(R)} r (x)$ tends to infinity, as
    $R$ does. Define 
    \begin{eqnarray}
    M(g,\Phi) & = & \lim_{R\to\infty} \frac 1 {16\pi} \int_{S(R)}
    (g_{ik,i}-g_{ii,k})\mathrm{d} S_k\;, \label{eq:mass_ADM}
    \\
    P_i(g,\Phi) & = & \lim_{R\to\infty} \frac 1 {8\pi} \int_{S(R)}
    P^{ij}\mathrm{d} S_j
    \end{eqnarray}
    (these
    integrals have to be calculated in the local $\alpha$--admissible
    coordinates $\Phi^i (p)=x^i$). $M$ and $P_i$ are finite,
    independent upon the particular family of spheres $S(R)$ chosen,
    provided $r(S(R))$ tends to infinity as $R$ does.
\end{theorem}

\begin{lemat}
    \label{lem:chr}
    Let $(g,\Phi_1)$ and $(g,\Phi_2)$ be
    $\alpha_1$ and $\alpha_2$--admissible, respectively, with any
    $\alpha_a
    > 0$. Let $\Phi_1 \circ\Phi_2^{-1}: \mathbb{R} ^3\backslash K_2 \rightarrow \mathbb{R} ^3\backslash K_1$
    be a twice differentiable diffeomorphism, for some compact sets
    $K_1$ and $K_2 \subset \mathbb{R}^3$. Then, in local coordinates
    $$\Phi^i_1 (p)=x^i\, , \qquad \Phi_2^i (p)=y^i\;,$$
    the diffeomorphisms $\Phi_1 \circ \Phi_2^{-1}$ and $\Phi_2 \circ
    \Phi_1^{-1}$ take the form
    $$x^i(y) = \omega^i{}_j\ y^i+\eta^i (y)\;,\qquad  y^i (x) = (\omega^{-1})^i{}_j\
    x^i\ + \zeta^i (x)\;,$$ $\zeta^i$ and $\eta^i$ satisfy, for some
    constant $C \in \mathbb{R}$,
    \begin{eqnarray*}
        &&|\zeta^i(x)| \le C (r(x)+1)^{1-\alpha}\;, \qquad
        |\zeta^i{}_{,j}(x)| \le C (r(x)+1)^{-\alpha}\;,
        \\ &&|\eta^i(y)| \le C (r(y)+1)^{1-\alpha}\;, \qquad |\eta^i{}_{,j}(y)|
        \le C (r(y)+1)^{-\alpha}\;,
        \\ && r(x) = (\sum (x^i)^2)^{1/2}\;,\qquad  r(y) = (\sum (y^i)^2)^{1/2}\;,
    \end{eqnarray*}
    with $\alpha=\min (\alpha_1, \alpha_2)\,, \, \omega^i{}_j$ is an $O(3)$
    matrix, and $r^0$ is to be understood as $\ln r$.
\end{lemat}
\begin{theorem}	
    Let $(g, \Phi_a), a=1,2$, satisfy the
    hypotheses of theorem \ref{thm:classical_ADM} and lemma \ref{lem:chr}. 
    Then
    \begin{enumerate}
        \item $M (g,\Phi_1)= M (g,\Phi_2)$
        \item $P_i (g,\Phi_1)= \omega_i{}^j P_j (g,\Phi_2)$ where $\omega \in O(3)$, given by lemma \ref{lem:chr}.
    \end{enumerate}
\end{theorem}
It seems unavoidable that a limiting process is involved in the definitions in theorem \ref{thm:classical_ADM}. But finding expressions that do not depend on the first derivatives but on rather more geometric quantities is an old question that has attracted the attention of many authors. It was suggested by Ashtekar and Hansen \cite{ashtekarhansen1978unified}, see also Chru\'sciel \cite{chrusciel1986remark}  that the mass could be rather defined from the Ricci tensor and a conformal Killing field of the Euclidean space. We now recall the alternative definition of asymptotic invariant via the Ricci tensor:
\begin{defi} \label{def:Ricci_mass}
    Let $X$ be the radial vector field $\mathcal{S}=r \partial_{r}$ in the chosen chart at infinity. Then, we define the Ricci version of the mass of $M$ by
    \begin{equation*}
    m_{R}(g)=-\frac{1}{8 \pi} \lim _{r \rightarrow \infty} \int_{S_{r}}\left({\overset{3}{R}{}^{k}}_{l }-\frac{1}{2} \overset{3}{R} {\delta^{k}}_{l } \right) \mathcal{S}^{l} \mathrm{d} S_{k}
    \end{equation*}
\end{defi}
Equality between the two definitions, as well as a similar identity for the center of mass, has then been proved rigorously by Huang using a density theorem \cite{huang2012center}, see also a simplified proof given by Herzlich in \cite{herzlich2016computing}.
\section{Lower order corrections for Ricci mass}
We consider a correction to the Ricci mass, definition \ref{def:Ricci_mass}, proportional to the curvature scalar
\begin{equation}
\tilde{m}_{R}=-\frac{1}{8 \pi} \int_{S(r)} \left(\overset{3}{R}{}^{i}{}_{j} -\frac12 \delta^{i}_{j} \overset{3}{R}\right)\mathcal{S}^j \mathrm{d} S_i -\frac{1}{8 \pi} \int_{S(r)}\left(\beta+\frac12\right) \delta^{i}_{j} \overset{3}{R} \, \mathcal{S}^j \mathrm{d} S_i
\label{eq:corrected_Ricci_mass}
\end{equation}
Using CKV equation and contracted Bianchi identity,
\begin{eqnarray}
\mathcal{S}_{(a|b)}&=&\lambda \gamma_{ab} \, ,\\
\left(\overset{3}{R}{}^{kl}-\frac{1}{2}\gamma^{kl} \overset{3}{R}\right)_{|k}&=&0 \, ,
\end{eqnarray}
we have
\begin{equation}
\left(\overset{3}{R}{}^{kl}\mathcal{S}_{l}+\beta \mathcal{S}^{k} \overset{3}{R}\right)_{|k}=\lambda \overset{3}{R}+\frac{1}{2}\overset{3}{R}_{|k}\mathcal{S}^{k}+\beta n \lambda \overset{3}{R}+\beta \overset{3}{R}_{|k} \mathcal{S}^{k}
\end{equation}
Assuming $\overset{3}{R}=p(\theta) r^{a}$ we have
\begin{eqnarray}
\left(\overset{3}{R}{}^{kl}\mathcal{S}_{l}+\beta \mathcal{S}^{k} \overset{3}{R}\right)_{|k}&=&\lambda \overset{3}{R}+\frac{1}{2}a \overset{3}{R}+\beta n \lambda \overset{3}{R}+\beta a \overset{3}{R}\\
&=&\overset{3}{R} \left(\lambda + \frac{1}{2} a+\beta(\lambda n+a)\right)
\end{eqnarray}
For example, we have calculated the corrected Ricci mass for Bowen--York spinning black hole. In this case, we have $\lambda=1, \, n=3, \, a=-6$. It leads to
\begin{equation}
\beta=-\frac{2}{3}
\end{equation}
Using constraint equation with $K=0$, $R-K^{kl}K_{kl}=0$ the following quantity is conserved up to controlled term
\begin{eqnarray}
-\frac{1}{8 \pi}\int_{S(R)}\left({R^{r}}_{e}-\frac{2}{3}K^{k l}K_{k l} \delta^{r}_{e}\right)\mathcal{S}^{e} \sqrt{\gamma}&=&m-\frac{4 J^{2}}{R^{3}}+\frac{2}{3} \frac{6 J^{2}}{R^{3}} +O\left(\frac{1}{R^4} \right) \nonumber\\
&=& m +O\left(\frac{1}{R^4} \right)
\end{eqnarray}
The correction enables on to get rid of unwanted term proportional to square of angular momentum. Similar mechanism can be observed for mass for asymptotically Kerr spacetimes \cite{smolka2017examination}.  
\section{Strong asymptotic flatness with the help of ACYK \label{sec:Strong_asymp_ACYK}}

Consider an asymptotically flat spacetime (at spatial infinity),
fulfilling the (complete non-linear) Einstein equations. Suppose,
moreover, that the energy--momentum tensor of the matter vanishes
around spatial infinity (``sources of compact support'').
This means that
the Riemann tensor and the Weyl tensor do coincide outside of the world
tube containing the matter. Let us analyze, for simplicity, this
situation in terms of an asymptotically flat coordinate system
(for nice geometric formulations of
asymptotic flatness see e.g. \cite{ashtekarhansen1978unified} or \cite{AR}).
We suppose that there exists an (asymptotically
Minkowskian) coordinate system $(x^\mu)$:
\[ g_{\mu\nu} - \eta_{\mu\nu} \sim r^{-b} \; \; \; \; \;
g_{\mu\nu,\lambda} \sim r^{-b -1} \] \noindent
where $\displaystyle r:= \sum_{k=1}^{3} (x^k)^2$ and typically $b=1$ (but
$1\geq b > \frac 12$ is also possible -- see \cite{Ch1}).

For a general asymptotically flat (AF) metric we cannot expect that the
equations ${\cal Q}_{\lambda \kappa \sigma}(Q,g)=0 \, ,$ see (\ref{eq:CYK_eq1}).
Instead, we assume that ${\cal Q}_{\lambda \kappa
    \sigma}(Q,g)$ has a certain asymptotic behavior at spatial infinity
\begin{equation}\label{QAF}
{\cal Q}_{\mu\nu\lambda} \sim r^{-c}  \end{equation}
On the other hand, suppose that  $Q_{\mu\nu}$ behaves asymptotically as
follows:
\[ Q_{\mu\nu} \sim r^{a} \; \; \; \; \;
Q_{\lambda \kappa , \sigma} \sim r^{a-1}  \]
\noindent
Moreover, suppose that the Riemann tensor $R_{\mu\nu\kappa\lambda}$
behaves asymptotically as follows:
\[ R_{\mu\nu\kappa\lambda} \sim r^{-b -1 -d} \] \noindent
It can be easily checked (see e.g. \cite{JNG}) that the vacuum Einstein
equations imply the following equality:
\begin{equation}\label{*R*}
\nabla_{\lambda} \left( {^*\!}R{^*}^{\mu\lambda}{_{\alpha\beta}}
Q^{\alpha\beta} \right) = - \frac 23
R^{\mu\lambda \alpha\beta} Q_{\lambda\alpha\beta} \end{equation}

The left--hand side of (\ref{*R*}) defines an asymptotic charge provided
that the right--hand side vanishes sufficiently fast at infinity. It is easy to
check that, for this purpose, the exponents $b,c,d$ have to fulfill the
inequality
\begin{equation}\label{2} b + c + d > 2  \end{equation}
In typical
situation when $b=d=1$, the inequality (\ref{2}) simply means that $c > 0$. In
this case a weaker condition is also possible (for example ${\cal
    Q}_{\mu\nu\lambda} \sim (\ln r)^{-1-\epsilon}$ with $\epsilon >0$).

Let us define an {\em asymptotic conformal Yano--Killing tensor} (ACYK) as an
antisymmetric
tensor $Q_{\mu\nu}$ such that ${\cal Q}_{\mu\nu\lambda} \rightarrow
0$ at spatial infinity.
For constructing the ACYK tensor we can begin with the CYK solutions in flat Minkowski space. Asymptotic behavior at infinity of these flat
solutions explain why we expect for any ACYK tensor the following
behavior:
\[ Q_{\mu\nu} = {^{(2)}\!}Q_{\mu\nu} + {^{(1)}\!}Q_{\mu\nu} +
{^{(0)}\!}Q_{\mu\nu} \] \noindent
where ${^{(2)}\!}Q_{\mu\nu} \sim r^2$, ${^{(1)}\!}Q_{\mu\nu} \sim r$ and
${^{(0)}\!}Q_{\mu\nu} \sim r^{1-c}$.

It is easy to verify that $c \geq b +1 -a$ and if $b=1$
than for $a=2$ we have $c \geq 0$. This means that in a general situation
there are no solutions of (\ref{QAF}) with nontrivial
${^{(2)}\!}Q_{\mu\nu}$ and $c>0$.
This is the origin of the difficulties with the definition of the angular
momentum.
On the other hand it is easy to check that the energy--momentum four--vector
and the dual one are well defined ($a=c=1$)  and the condition
$b + d > 1$ can be easily fulfilled (typically $b=d=1$).

\paragraph{Strong asymptotic flatness} Here, we propose a new, stronger definition of the asymptotic flatness.
The definition is motivated by the above discussion.

Suppose that there exists a coordinate system $(x^\mu)$
such that:
\[ g_{\mu\nu} - \eta_{\mu\nu} \sim r^{-1} \]
\[ \Gamma^{\kappa}{_{\mu\nu}} \sim r^{-2} \]
\[ R_{\mu\nu\kappa\lambda} \sim r^{-3} \] \noindent
In the space of ACYK tensors fulfilling the asymptotic condition
\begin{equation}\label{Q1}
Q_{\lambda (\kappa ;\sigma)}
-Q_{\kappa (\lambda ;\sigma)} +
g_{\sigma[\lambda} Q_{\kappa ]}{^\delta}_{;\delta} =
{\cal Q}_{\lambda\kappa\sigma} \sim r^{-1}  \end{equation}
we define the following equivalence relation:
\begin{equation}\label{rel} Q_{\mu\nu} \equiv Q_{\mu\nu}'
\Longleftrightarrow Q_{\mu\nu}  - Q_{\mu\nu}' =  O(1)
\end{equation}
for $r \rightarrow \infty$.
We assume that the space of equivalence classes defined by (\ref{Q1}) and
(\ref{rel}) has a finite dimension $D$ as a vector space.
The maximal dimension $D = 14$ correspond to the situation where there
are no supertranslation problems
in the definition of an angular momentum.
In the case of spacetimes for which $D<14$
the lack of certain ACYK tensor means that the corresponding charge
is not well defined.
\section*{Data availability statement}
The paper has no associated data. All concepts and logical implications are given in the manuscript.
\bibliography{BibTex_asymp_charg}
\end{document}